\documentclass[a4paper,11pt]{article}
\usepackage{amsmath}
\usepackage{amssymb}
\usepackage{graphicx}

\usepackage[utf8]{inputenc} 

\usepackage[square,numbers,comma,sort&compress]{natbib}
\usepackage[colorlinks=true,citecolor=blue,linkcolor=purple,urlcolor=purple]{hyperref}
\usepackage{geometry}
\usepackage[usenames,dvipsnames,svgnames,table]{xcolor}
\usepackage{booktabs}
\usepackage{slashed}

\def\varabstract{ }
\def\varkeywords{ }
\def\vararxivnumber{ }
\def\vartitle{ }
\def\varsubtitle{ }
\renewcommand{\title}[1]{\gdef\vartitle{#1}}

\renewcommand{\abstract}[1]{\gdef\varabstract{#1}}
\newcommand{\keywords}[1]{\gdef\varkeywords{#1}}

\newtoks\authtoks
\renewcommand{\author}[2][]{%
	\authtoks=\expandafter{\the\authtoks#2$^{#1}$\ }%
}
\newtoks\affiltoks
\newcommand{\affiliation}[2][]{%
    \affiltoks=\expandafter{\the\affiltoks{\item[$^{#1}$]#2}}%
}
\newtoks\emailtoks\newcounter{emailcounter}%
\newcommand{\emailAdd}[1]{%
\ifnum\theemailcounter>0\emailtoks=\expandafter{\the\emailtoks, \typeemail{#1}}%
\else\emailtoks=\expandafter{\typeemail{#1}}%
\fi
\stepcounter{emailcounter}}
\newcommand{\typeemail}[1]{\href{mailto:#1}{\tt #1}}

\renewcommand\maketitle{
	\newgeometry{margin=2cm}
	\pagestyle{empty}\setcounter{page}{0}
	{\huge\flushleft\sffamily\bfseries\vartitle\\\Large\varsubtitle\par}
\vskip6ex
{\large\bfseries\raggedright\sffamily\the\authtoks\par}
\vskip2ex
\begin{list}{}{%
\setlength{\leftmargin}{0.28cm}%
\setlength{\labelsep}{0pt}%
\setlength{\itemsep}{-3pt}%
\setlength{\topsep}{-\parskip}}
\itshape\small%
\the\affiltoks
\end{list}
\vskip2ex
\noindent\hspace{0.28cm}\begin{minipage}[l]{.9\textwidth}
\begin{flushleft}
\textit{E-mail:} \the\emailtoks
\end{flushleft}
\end{minipage}
\vskip5ex
\noindent{\renewcommand\baselinestretch{.9}\textsc{Abstract:}}\ \varabstract
\vskip5ex 
\if!\varkeywords!\else\noindent{\textsc{Keywords:}}\ \varkeywords \vskip2ex\fi
\if!\vararxivnumber!\else\noindent{\textsc{ArXiv ePrint:}} \href{http://arxiv.org/abs/\vararxivnumber}{\vararxivnumber}\vskip2ex\fi

\newpage
\restoregeometry
\pagestyle{plain}

\setcounter{footnote}{0}
} 

\usepackage[square,numbers,comma,sort&compress]{natbib}

\definecolor{MS}{rgb}{0,0,1}

\newcommand{\blX}{\mbox{\boldmath {\bf X}}}
\newcommand{\blA}{\mbox{\boldmath {\bf A}}}

\newcommand{\blB}{\mbox{\boldmath {\bf B}}}
\newcommand{\blV}{\mbox{\boldmath {\bf V}}}
\newcommand{\blb}{\mbox{\boldmath {\bf b}}}

\title{Dynamical projections for the visualization of PDFSense data}

\author[1]{Dianne Cook,}\emailAdd{dicook@monash.edu}
\author[1,2]{Ursula Laa,}\emailAdd{ursula.laa@monash.edu}
\author[2]{and German Valencia}\emailAdd{german.valencia@monash.edu}

\affiliation[1]{School of Econometrics and Business Statistics, Monash University,
Melbourne VIC-3800}

\affiliation[2]{School of Physics and Astronomy, Monash University,
Melbourne VIC-3800}

\abstract{A recent paper on visualizing the sensitivity of hadronic experiments to nucleon structure~\cite{Wang:2018heo} introduces the tool 
PDFSense which defines measures to allow the user to judge the sensitivity of PDF fits to a given experiment. The sensitivity is characterized by high-dimensional data residuals that are visualized in a 3-d subspace of the 10 first principal components or using t-SNE~\cite{vanDerMaaten2008}. We show how a tour,  a dynamic visualisation of  high dimensional data, can extend this tool beyond 3-d relationships. This approach enables resolving structure orthogonal to the 2-d viewing plane used so far, and hence finer tuned assessment of the sensitivity. }


\begin{document}

\maketitle

\section{Introduction}

Many problems in physics can be broadly characterized as a description of a large number of observations with models that contain multiple parameters. It is common practice to perform a  global fit to the observations to arrive at the set of parameter values that best fits the data. To understand how well this fit describes the observations, a series of one or two-dimensional projections of confidence level regions are usually provided. 

It is desirable to visually inspect the results of such fits to gain insight into their structure. One possibility is to directly compare the predictions of different parameter sets in the vicinity of the best fit. A simple algorithm to organise this idea that results in a manageable number of such parameter sets can be constructed using singular value decomposition (SVD). One first decides the confidence level at which to make the desired comparison and quantifies it with the corresponding $\Delta\chi^2$  for the appropriate number of parameters being fit, $n$. The region in parameter space within the desired confidence level is approximately an $n$-dimensional ellipsoid, and SVD provides an ideal set of 2$\times n$ 
points on which to evaluate the predictions of the model for visual inspection. These points are given by the intersections of the ellipsoid with its principal axes and clearly provide a minimal sample of parameter space that covers all relevant directions at a desired confidence level. 

A tool for the direct visualisation of the high dimensional model predictions thus constructed  has existed in the statistics literature for many years, but has not been applied to high energy physics problems recently.\footnote{The precursor~\cite{FFT74v} of this tool was originally developed to tackle problems in high energy physics.} It is called a tour, and is a dynamic visualization of low-dimensional projections of high-dimensional spaces. The most recent incarnation of the tool is available in the R~\cite{citeR} package, called {\tt tourr}~\cite{tourr}. The goal of this paper is to introduce the use of a tour as a visualisation tool for sensitivity studies of parton distribution functions (PDFs) building on the formalism that has been developed over the years by the CTEQ collaboration. It is beyond the scope of this article to provide a detailed analysis of the PDF uncertainties. The choice of this example has two motivations: the PDF fits embody the generic problem of multidimensional fits to large numbers of observables that are common in high energy physics; and Ref.~\cite{Wang:2018heo} has recently provided the parameter sets for this problem in an initial effort to visualize the PDF fits. Our starting point will be the PDFSense \cite{Wang:2018heo} results but our study differs in an important way: PDFSense 
utilizes the Tensorflow Embedding Projector~\cite{tfep}, 
limiting visualisation to three of the first ten principal components, that is, a 3-d subspace, whereas the tour allows us to explore the full space. As we will see here, this allows additional insights into the fits. 

Our paper is organised as follows. In Section~\ref{sec:residuals} we first describe the problem as formulated in Ref.~\cite{Wang:2018heo} and we discuss a toy example to illustrate the concepts involved. 
We then 
introduce the tour algorithm and its implementation in Section~\ref{sec:Tour}. Finally we  discuss the results obtained by applying tour to the PDFSense dataset in Section~\ref{sec:Results} and present our conclusions in Section~\ref{sec:Conclusions}.

\section{PDF fits and residuals}
\label{sec:residuals}

The analysis of collider physics results relies on theoretical calculations of cross-sections and distributions. Factorization theorems allow us to bypass non-perturbative physics that cannot be calculated from first principles and to describe instead, the initial state of a reaction in terms of parton distribution functions or PDFs. These consist of simple functional forms describing the probability density for finding a given quark or gluon in the proton  with a given momentum fraction $x$,  at a given momentum transfer scale $Q$. The PDFs used today have been constructed by fitting high energy physics data collected over many years by multiple experiments and are produced by large collaborations. As such, they constitute an ideal example of a multidimensional parameter fit to a large data set to study with a tour.

For our study we will make use of the framework for treating uncertainties of the PDF predictions as has been defined in~\cite{Pumplin:2000vx,Pumplin:2001ct}. The best fit PDF, defined by the set of $n$ parameters $a^0_i$, is obtained by finding the global minimum of a $\chi^2$ function. To study uncertainties in the fit one considers small variations of the parameters around the minimum using a quadratic approximation for the $\chi^2$ function written in terms of the Hessian matrix of second derivatives at the minimum, H. The eigenvectors of this matrix provide the principal axes of the confidence level ellipsoids around the global minimum, and one defines a displacement along these directions to find the $n$ dimensional set of points $a_i$ which provide $2n$ PDF sets that differ from the best fit by a desired confidence level. 

Ref. \cite{Wang:2018heo} has introduced the package PDFSense to study the sensitivity of different experiments to different aspects of the PDFs. An ingredient of that study are the so-called shifted residuals which are related to the experimental error contribution to the $\chi^2$ by~\cite{Stump:2001gu}
\begin{equation}
\chi^2_E (\vec{a}) = \sum_{i=1}^{N_d} r^2_i(\vec{a}) + \sum_{\alpha=1}^{N_{\lambda}}\bar{\lambda}_{\alpha}^2(\vec{a})
\end{equation}
where the $\bar{\lambda}_{\alpha}$ are the best-fit nuisance parameters.
The shifted residuals $r_i(\vec{a})$ are calculated as the difference between the theoretical prediction $T_i(\vec{a})$
and the shifted central data value $D_{i,sh}(\vec{a})$, normalised by the total uncorrelated uncertainty $s_i$,
\begin{equation}
r_i(\vec{a}) = \frac{1}{s_i}(T_i(\vec{a}) - D_{i,sh}(\vec{a})).
\end{equation}
Note that $D_{i,sh}(\vec{a})$ is the observed central value shifted by a function of the optimal nuisance parameters $\bar{\lambda}_{\alpha}$ and
therefore depends on the point in parameter space considered.
The so-called response of a residual to an experimental result $i$ is then defined as~\cite{Wang:2018heo}
\begin{equation}
\delta_{i,l}^{\pm} \equiv (r_i(\vec{a}_l^{\pm})-r_i(\vec{a}_0))/\langle r_0\rangle_E
\end{equation}
with $\langle r_0\rangle_E$ the root-mean squared residuals characterizing the quality of fit to experiment $E$.\footnote{Note
that the shifted central data value enters the residuals, thus while the observed central value cancels in the definition of $\delta_{i,l}^{\pm}$,
differences in the shift arising from differences of the optimized nuisance parameters at $\vec{a}_l^{\pm}$ are encoded in the results together with difference in theory predictions.}
It parameterizes the change in residuals with variations along the independent directions  $\vec{a}_l^{\pm}$.
Large values of $\delta_{i,l}^{\pm}$ therefore indicate considerable variation in the theory prediction values within the selected window of allowed probability variation
along the considered direction.
We thus consider a $2N$ dimensional vector
\begin{equation}
\vec{\delta}_i=\{\delta_{i,1}^{+},\delta_{i,1}^{-},...,\delta_{i,N}^{+},\delta_{i,N}^{-}\}.
\end{equation}
for each data point (i.e.\ experimental result).
Concretely, here we consider a 56 dimensional parameter space in which we want to compare and group the experimental results. These responses $\vec{\delta}_i$ are calculated and provided by Ref.~\cite{Wang:2018heo} and they constitute the starting point of our study.

\subsection{Simple illustrative example}
\begin{figure}[ht!]\centering
\includegraphics[width=.45\textwidth]{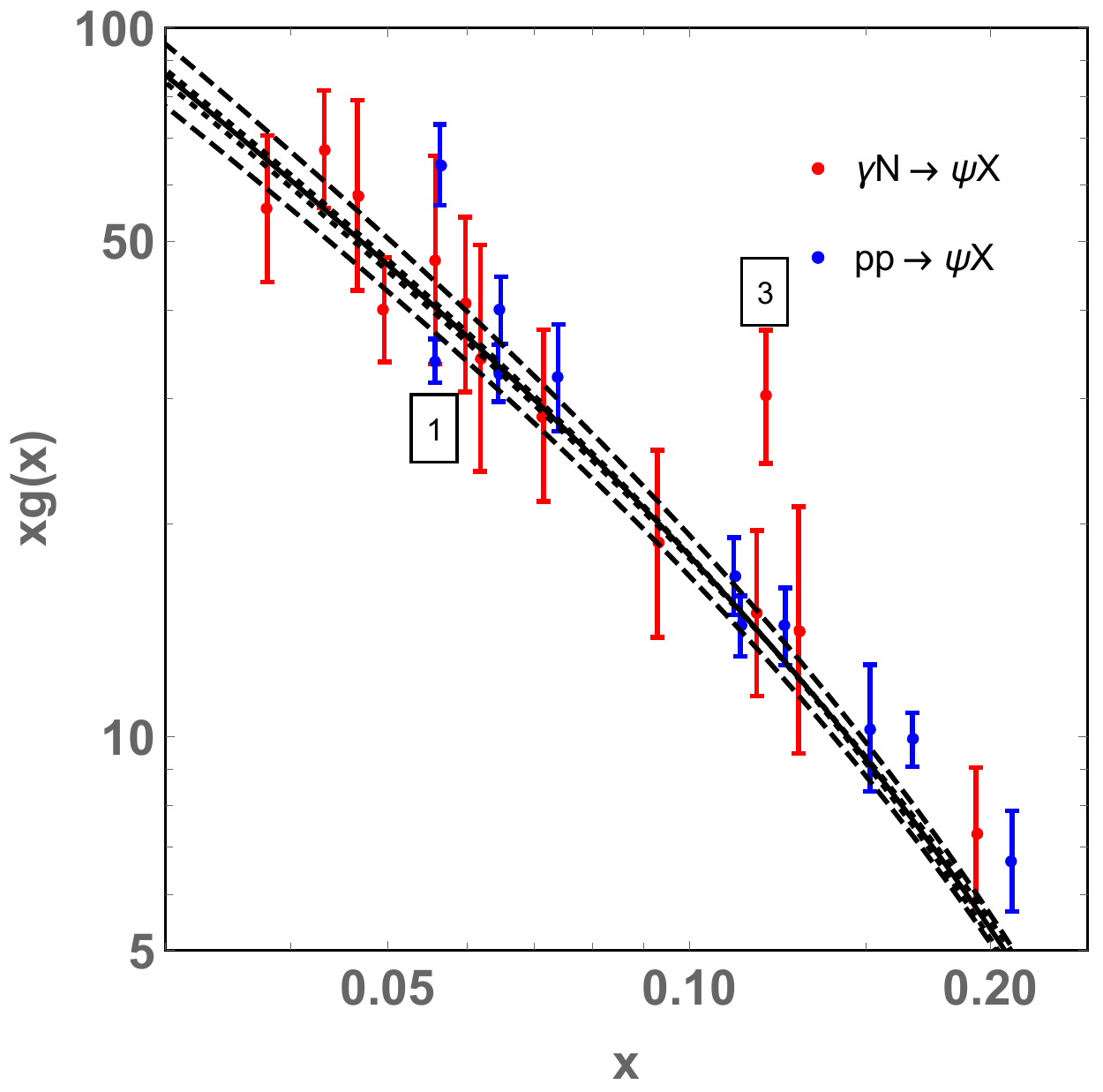}\includegraphics[width=.45\textwidth]{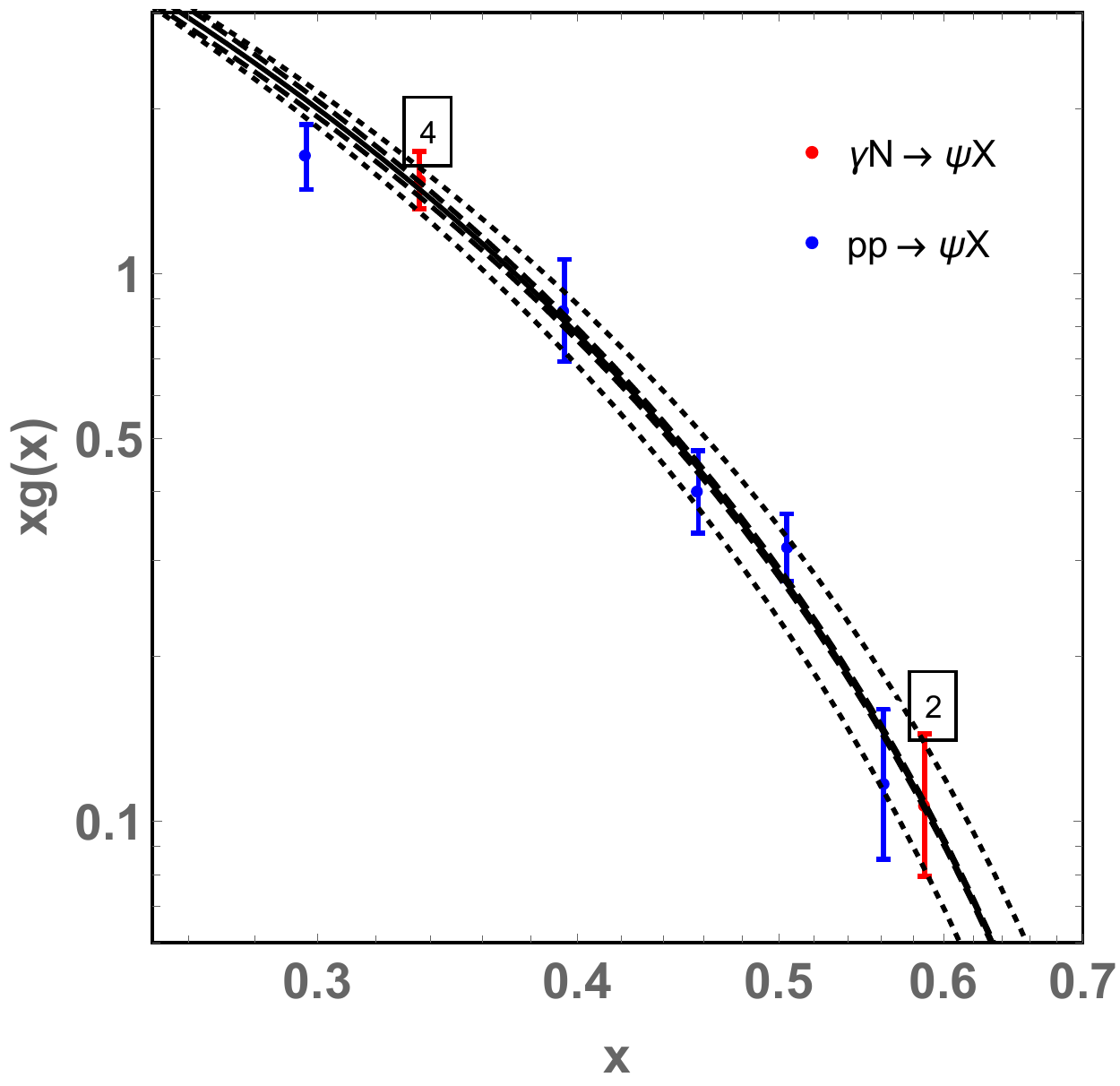}
\caption{For illustrative purposes, two data sets of gluon parton distribution function, in the form $p(x)\pm \Delta p(x)$ for 15 and 16 values of $x$, respectively (shown in red and blue). The left (right) panel shows the low (high) $x$ region respectively.
\label{fig:gdata}}
\end{figure}

\begin{figure}[ht!]\centering
\includegraphics[width=0.4\textwidth]{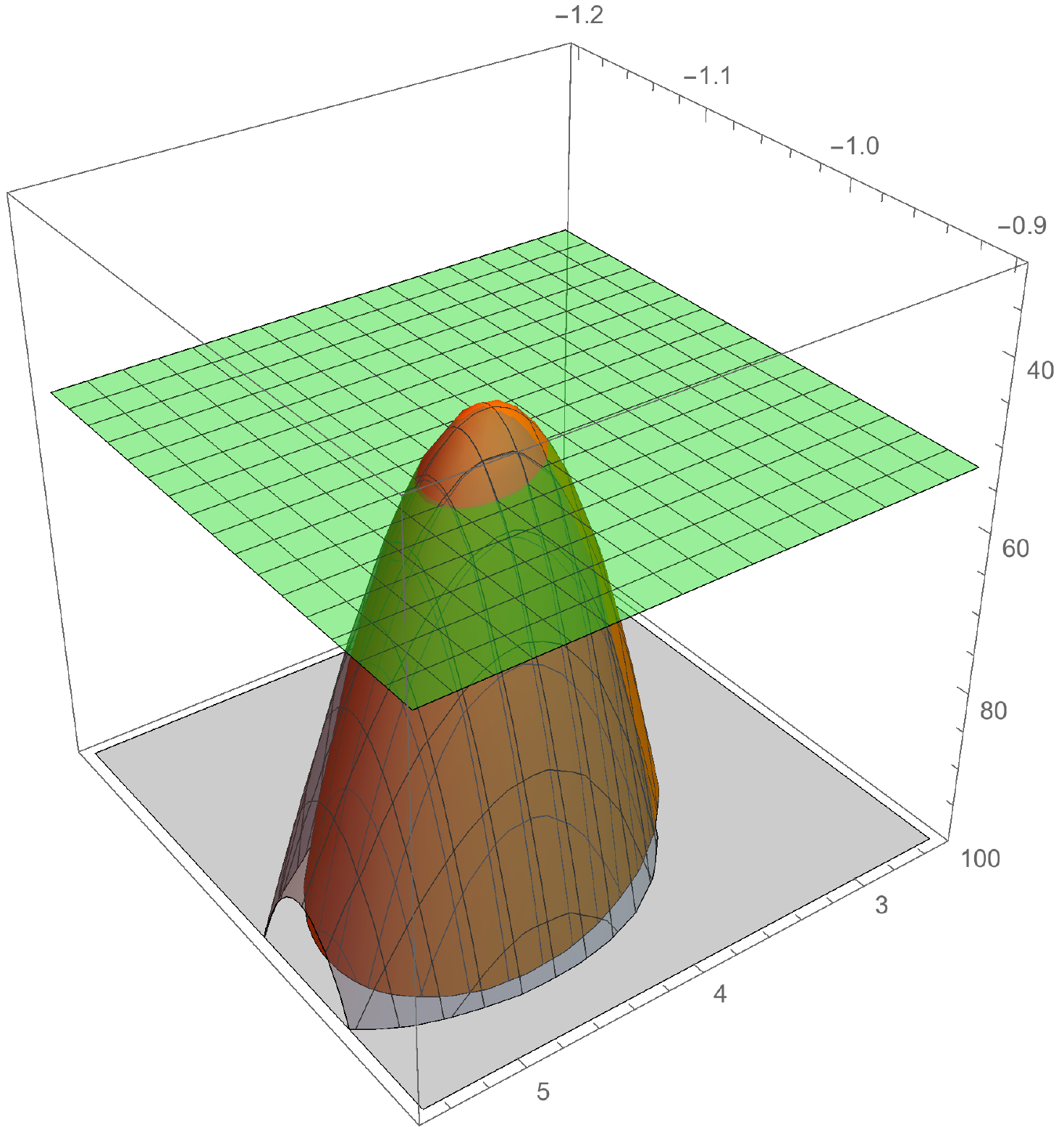}\includegraphics[width=0.4\textwidth]{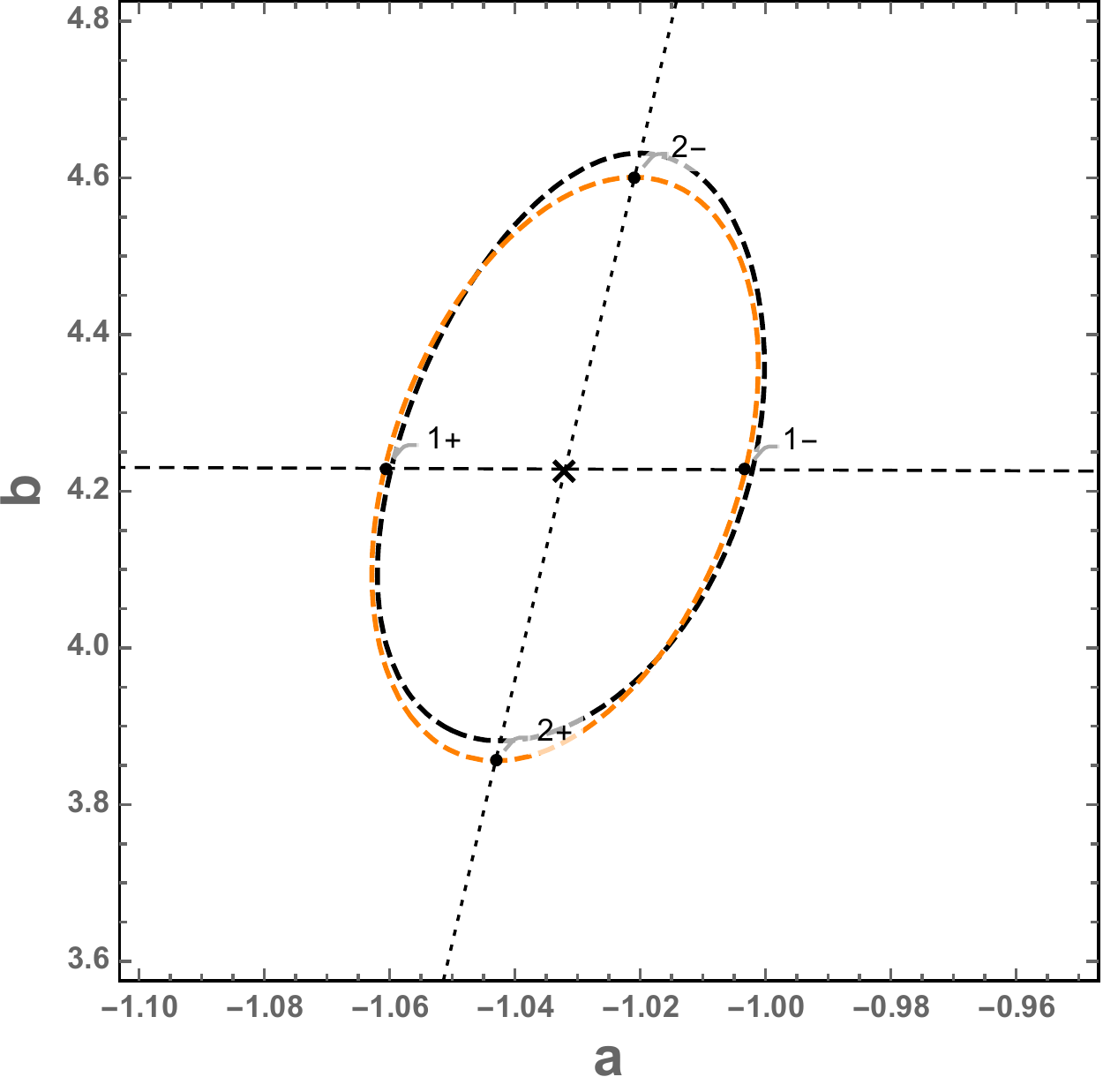}
\caption{Difference between the $\chi^2$-function (black), and quadratic approximation (orange). Their intersection with a 95\% confidence level plane is shown on the right panel. The intersections of the principal axes  with the ellipse (that occurs in the quadratic approximation) are shown as the black dots in the right panel. The numbers label the eigenvector of H corresponding to that direction.
\label{fig:95pc}}
\end{figure}

The procedure described so far has been used for many years, but it is complicated. For newcomers to the field, we illustrate it here using a simple example drawn from  two early data sets for the gluon parton distribution function extracted from two types of $\psi$ production experiments \cite{Barger:1979js}. This example will allow us to illustrate all the concepts involved. In Figure~\ref{fig:gdata} we show these two data sets, labelling the points and their error bars $p(x)\pm \Delta p(x)$, for 15 and 16 values of $x$ (in red and blue)  respectively. The points are  fit to the two-parameter function
\begin{eqnarray}
g(a,b,x)=\frac{1}{2}(1+b)(1-x)^b x^a,
\label{eq:pdfEx}
\end{eqnarray}
similar to but simpler than the forms used today. The next step is to minimise the $\chi^2$-function defined by
\begin{eqnarray}
\chi^2(a,b)=\sum_{x_i}\left(\frac{g(a,b,x_i)-p(x_i)}{\Delta p(x_i)}\right)^2.
\end{eqnarray}
The parameters $a_0,b_0$ that result in the global minimum $\chi^2(a,b)_{\rm min}$ define the best fit to the data. They are shown as the cross in the right panel of Figure~\ref{fig:95pc},  and produce 
the solid black curve shown in Figure~\ref{fig:gdata}.  At the same time one adopts a quadratic approximation to the $\chi^2$ function in the vicinity of its minimum 
\begin{eqnarray}
\chi^2(a,b)\approx \chi^2(a_0,b_0) +\frac{1}{2}
\left(
\begin{array}{cc}
 a-a_0 &   b-b_0
\end{array}
\right)
\left(
\begin{array}{cc}
\frac{\partial \chi^2(a,b)}{\partial a^2}  &  \frac{\partial \chi^2(a,b)}{\partial a\partial b}      \\
 \frac{\partial \chi^2(a,b)}{\partial a\partial b}   &     \frac{\partial \chi^2(a,b)}{\partial b^2}  
\end{array}
\right)_0 
\left(
\begin{array}{c}
   a-a_0   \\
 b-b_0
\end{array}
\right),
\end{eqnarray}
where the matrix of second derivatives evaluated at the global minimum is the well-known Hessian. This approximation seems unnecessary for the simple example we are discussing now but is used for the current global fits offering complementary features to exact numerical methods \cite{Hou:2016sho}.
To quantify the error in the fit one then constructs the region in $a,b$ parameter space corresponding to a given confidence level. For our example we take  $\chi^2(a,b)-\chi^2(a_0,b_0)\leq 5.99$ which corresponds to a 95\% confidence level in the estimation of two parameters. The intersection of the plane $\chi^2(a,b)=\chi^2(a_0,b_0)+ 5.99$ (green) with the $\chi^2(a,b)$ function (shown in black) and its quadratic approximation (in orange) is shown in the left panel of Figure~\ref{fig:95pc}. The right panel in the same figure shows the ellipsoid (two-dimensional in this case) defined by this intersection for the quadratic approximation (in orange) and the deformed ellipsoid in black for the exact $\chi^2(a,b)$ function. The difference between the two is small indicating that the quadratic approximation is quite adequate for this confidence level. 
The eigenvectors of the Hessian matrix provide the directions of the principal axes of the ellipsoid and are shown in black in the right panel of Figure~\ref{fig:95pc}: the dashed (dotted) lines correspond to the direction associated with the largest (smallest) eigenvalue.  The intersections of these axes with the ellipse, shown as  black dots, provide a set of  fits to the data that can be compared with the best fit and used as a means of quantifying the uncertainty in the fitting procedure. These are also shown in Figure~\ref{fig:gdata}.
 
\begin{figure}[ht!]\centering
\includegraphics[width=0.5\textwidth]{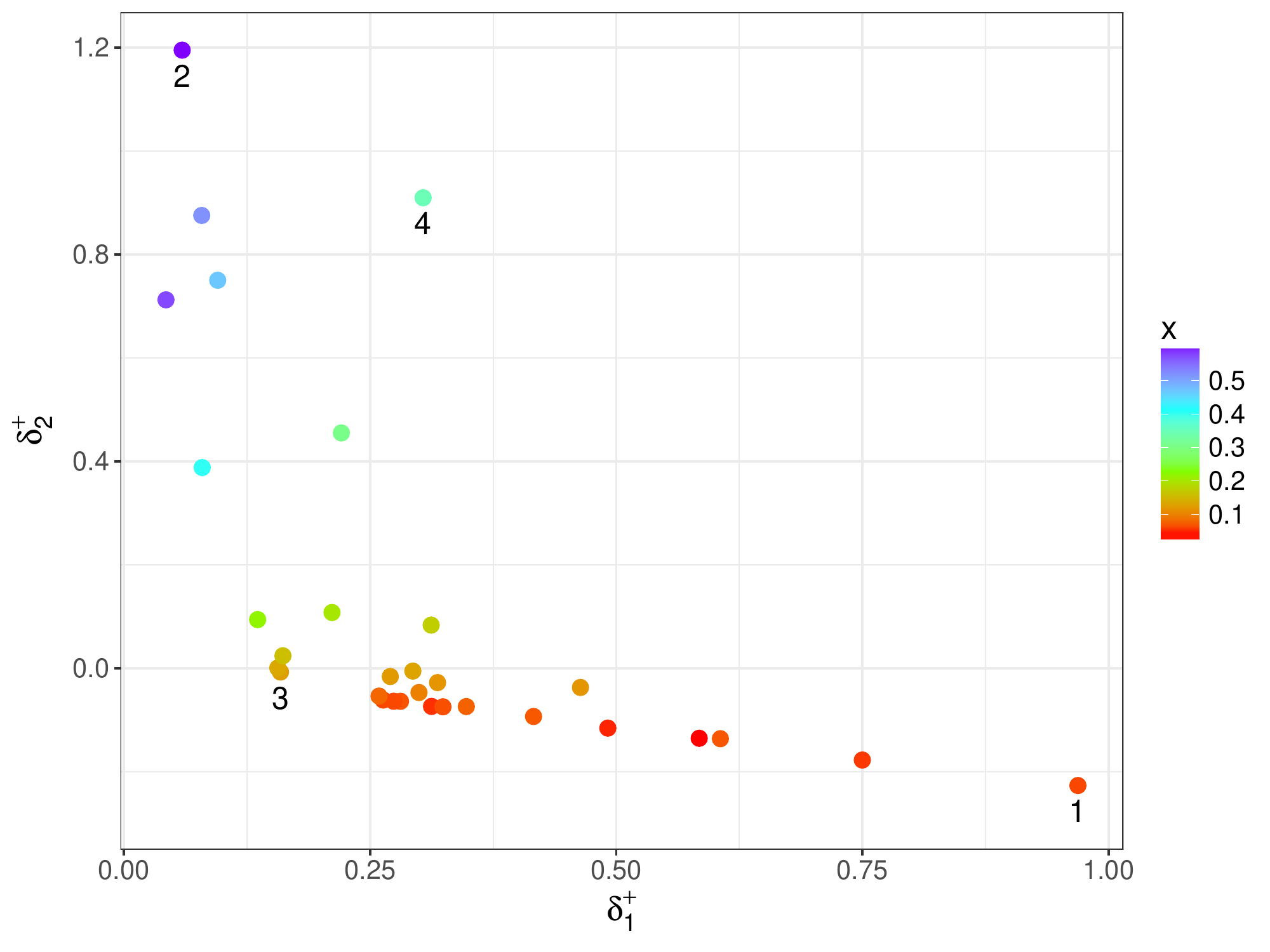}
\caption{The $\delta$ parameter space of the simple illustrative example: $\delta_i^+$ form the axes and color indicates the respective value of $x$. Note that only $\delta_i^+$ is shown because for this problem the $\delta_i^-$ directions contain the same information. Labelled points are the same as those labelled in Figure~\ref{fig:gdata}, and illustrate key features of the fits. \label{fig:deltas}}
\end{figure}

The set of responses, $\delta_{i,l}^{\pm}$, in this example is shown in Figure~\ref{fig:deltas}. From inspecting the limiting behaviour of Eq.~\ref{eq:pdfEx} it is clear that the description at low $x$ is dependent mainly on $a$ while large values of $x$ are mostly sensitive to $b$. This is reflected in the uncertainty curves in Figure~\ref{fig:gdata}, and also when looking at the $\delta$s. For this simple example the main directions identified by the Hessian method are in fact well aligned with the original directions in parameter space. Considering the values of $\delta$ we find that $\delta_1^{\pm}$, which corresponds mainly to a variation of $a$, takes large values for bins with low values of $x$, while $\delta_2^{\pm}$ takes large values for bins with large values of $x$. We conclude that the parameter dependence is captured by the $\delta$s as expected. Going to more complex descriptions and fits, as we do in the following,  this correspondence is no longer clear from the description and the $\delta$ values may be used to infer the parameter dependence of a given prediction.

In Figures~\ref{fig:gdata}~and~\ref{fig:deltas} we have labelled the following four points:
\begin{enumerate} \itemsep -0.07in
\item  point with highest value in $\delta_1$, found at low $x$ and with small error bar
\item  point with parametrized highest value in $\delta_2$, also has the highest value of $x$
\item  point that is not well described by the fits, but  has small values of $\delta$
\item  point with intermediate value of $x$ and small errors result in larger values in both $\delta$ directions.
\end{enumerate}
These observations illustrate that large values of $\delta$ correlate with points with errors that are comparable to or smaller than the uncertainty in the fit as parametrized by the Hessian method. At the same time, points that are not well described by the fits do not necessarily result in large $\delta$s.

\section{Data visualisation}
\label{sec:Tour}

When looking for structure in high dimensional parameter spaces we rely on tools for dimensional reduction and visualisation.
Due to the importance of this task, many methods have been developed. Here we give a brief overview of the tools used in the following work. 

\subsection{Dimension reduction} 
\subsubsection{Principal component analysis}
Principal component analysis (PCA) is an orthogonal linear transformation of elliptical data into a coordinate system, such that the first basis aligns with the maximum variance. The second basis is the direction of maximum variation orthogonal to the first coordinate, and the remaining bases are sequentially computed analogously. It is typically used for dimension reduction. To choose the number of principal components (PCs) to use, the proportion of variance explained by each component is examined. Either a pre-determined proportion of total variance is used, or by plotting the proportions against the number of PCs and choosing the point where this flattens to zero.

PCA is an optimization problem with a well defined solution. However, the outcome of the PCA is affected by the preparation of the input data.  The preparation can also be used to highlight specific aspects of the data distribution. For example, the input data is generally centered before performing PCA  by setting each variable to have a mean value of zero. In this way, large variation describing only mean values different from zero are removed from the results. Another approach would be to normalize the distribution, to emphasize directional information. Typically this means ``sphering'' of the data points, by normalizing each vector to have length one. This results in comparison of similar, or different, directions in the parameter space, but information about the differences in length are lost by this approach.

In this work we use the standard implementation {\tt prcomp} in R for the computation of the principal components.

\subsubsection{Nonlinear embeddings}
It is also common to examine non-linear mapping of the data points onto a low dimensional embedding. The aim is to preserve multidimensional structure by minimizing the difference in distances in the full parameter space as compared to distances in the low dimensional projection. PCA is a simple member of this more general type of transformation. A widely used method in machine learning is the algorithm called t-distributed stochastic neighbor embedding (t-SNE)~\cite{vanDerMaaten2008}. It has a goal to cluster similar points together (i.e. points with small Euclidean distance) while separating the individual clusters from one another. This gives appealing and often useful pictures but results should be considered with care as t-SNE is a nonlinear transformation and does not preserve original distance.
Note that while nonlinear embeddings may be useful in identifying clusters in the data, their interpretation is limited by lack of an analytical description of the transformation. This is not the case for linear transformations such as the PCA, where the transformation can be readily reversed to identify the contribution of the original parameters to a given principal component direction.

\subsection{Tour algorithm}

\subsubsection{Overview}

When a data set has more than two parameters, the tour~\cite{Asimov:1985:GTT:2812.2906} can be used to plot the multiple dimensions. Currently the typical approach is to plot two parameters or pairs of combinations of the parameters. The tour extends this idea to plot all possible combinations. The viewer is  provided with a continuous movie of smooth transitions from one combination to another, from which it is possible to extrapolate the shape of the parameter space in high-dimensions. Seeing many combinations in quick succession shows the associations between all the parameters. 

There are several types of tours. Here we use a grand tour, of projections from $n$-dimensional parameter space to 2-d projections space. A projection of data is computed by multiplying an $ m \times n$ data matrix, $\blX$, having $m$ sample points in $n$ dimensions, by an orthonormal $n \times d$ projection matrix, $\blA$, yielding a $d$-dimensional projection.  The grand tour is a mechanism for choosing which projections to display, and how the smooth transitions happen. New projections are chosen from all possible projections, and a geodesic interpolation to a target projection provides the smooth transition. The original algorithm is documented in \cite{Buja:2004}. The implementation used in this paper is from the {\tt tourr}~\cite{tourr} package in R~\cite{citeR}. 

The tour shows linear projections of the parameter space. In contrast, methods like t-SNE~\cite{vanDerMaaten2008} produce non-linear mappings from high- to low- dimensional space. The difference is that the shape of the data in high-dimensions is preserved by linear projections, but not with nonlinear mappings. 

\subsubsection{Algorithm}

A movie of data projections is created by interpolating along a geodesic path from the current (starting) plane
to the new target plane. In the grand tour, the target plane is chosen by randomly selecting a plane. The interpolation algorithm (as described in~\cite{Cook2008}) follows these steps:

\begin{enumerate}  \itemsep 0in
\item Given a starting $n\times d$ projection $\blA_a$, describing the
starting plane, create a new target projection $\blA_z$, describing
the target plane. It is important to check that $\blA_a$ and $\blA_z$
describe different planes, and generate a new $\blA_z$ if necessary. To
find the optimal rotation of the starting plane into the target plane
we need to find the frames in each plane which are the closest.
\item Determine the shortest path between frames using singular value
decomposition. $\blA_a'\blA_z=\blV_a\Lambda\blV_z',
~~~\Lambda=\mbox{diag}(\lambda_1\geq\dots\geq\lambda_d)$, and the
principal directions in each plane are $\blB_a=\blA_a\blV_a,
\blB_z=\blA_z\blV_z$, a within-plane rotation of the descriptive bases
$\blA_a, \blA_z$ respectively. The principal directions are the frames
describing the starting and target planes which have the shortest
distance between them. The rotation is defined with respect to these
principal directions. The singular values, $\lambda_i, i=1,\dots, d$,
define the smallest angles between the principal directions.
\item Orthonormalize $\blB_z$ on $\blB_a$, giving $\blB_*$, to
create a rotation framework.
\item Calculate the principal angles, $\tau_i =
\cos^{-1}\lambda_i, i=1,\dots, d$. 
\item Rotate the frames by dividing the angles into increments,
$\tau_i(t)$, for $t\in(0,1]$, and create the $i^{th}$ column of the
new frame, $\blb_i$, from the $i^{th}$ columns of $\blB_a$ and
$\blB_*$, by $\blb_i(t) = \cos(\tau_i(t))\blb_{ai} +
\sin(\tau_i(t))\blb_{*i}$. When $t=1$, the frame will be $\blB_z$.
\item Project the data into $\blA(t)=\blB(t)\blV_a'$.
\item Continue the rotation until $t=1$. Set the current projection to
be $\blA_a$ and go back to step 1.
\end{enumerate}

In a grand tour the target plane is drawn randomly from all possible target planes, which means that any plane is equally likely to be shown. That is, we are sampling from a uniform distribution on a sphere. To achieve this, sample $n$ values from a standard univariate normal distribution, resulting in a sample from a standard multivariate normal. Standardize this vector to have length equal to one, gives a random value from a $(n-1)$-dimensional sphere, that is, a randomly generated projection vector. Do this twice to get a 2-dimensional projection, where the second vector is orthonormalized on the first. 

The data typically needs some standardization or scaling before computing the tour. This can be as simple as centering each variable on 0, and standardizing to a range of -1 to 1. It could be as severe as sphering the data which in statistics means that the data is transformed into principal components (from elliptical shape to spherical shape). The same term is used for a different type of transformation in other fields, where observations are scaled to fall on a high-dimensional sphere, by scaling each observation to have length 1. (An interesting diversion: this type of sphering is the same transformation made on multivariate normal vectors to obtain a point on a sphere, to choose the target planes in the grand tour.)   

The initial description of the tour  promised display of all possible projections. Theoretically this is true, but practically it would require that the user stay watching forever! However, the coverage of the space is fairly fast, depending on $n$, and within a short time it is possible to guarantee all possible projections are displayed within an angle of tolerance.

\subsubsection{Display}

For physics problems, setting $d=2$ would be most common. The projected data is displayed as a scatterplot of points. It is also possible to overlay confidence regions, or contours. Groups in the data can be highlighted by color. Displaying the combination of variables of a particular projection can be useful to interpret patterns. This can be realized by plotting a circle with segments indicating the magnitude and direction of the contribution, and it is called the axes.  

The same tour path can be used to display subsets of the data, in different plots, to compare groups. When we break the display into subsets, the full data is also shown in each plot, in light grey. This makes it easier to do group comparison. 

\section{Results}
\label{sec:Results}

This section compares the findings made using the tour relative to those made with PDFSense using the recent CT14HERA2 fits~\cite{Hou:2016nqm}. The PDFSense results form the basis on which to expand the knowledge of PDF fits. 
The results from both tools are summarized in Table~\ref{tab:comparison}, where PDFSense results were obtained using the TensorFlow Embedding Projector (TFEP) software~\cite{tfep} for the visualisation of high-dimensional data.
The summary statistic ``reciprocated distance'' referenced in Table~\ref{tab:comparison} is defined as:
\begin{equation}
\mathcal{D}_{i}\ \equiv\ \left(\sum_{j\neq i}^{N_{\mathit{all}}}\frac{1}{|\vec{\delta}_{j}-\vec{\delta}_{i}|}\right)^{-1}.
\label{eq:recDist}
\end{equation}
TFEP provides two methods, PCA and t-SNE, and~\cite{Wang:2018heo} is exploring both for the visualisation of the data set. The PCA implementation returns projections onto the 10 first PCs evaluated from centered and sphered data, and allows the user to choose two or three of them to view the results. 

\begin{table}[htp]
\makebox[\textwidth]{\begin{tabular}{|c|p{6cm}|p{6cm}|}
\hline
 & PDFSense \& TFEP & Tour\\
\hline
1 &
Three clusters can be separated in the visualisation, labelled DIS, VBP and jet cluster. In the selected view the jet cluster is roughly orthogonal to the DIS cluster. &
We observe the differences in distributions between the three clusters more clearly. Substructure within the clusters is also observed, and studied in some detail.\\
\hline
2 & New ATLAS and CMS results will dominate the jet cluster. &
A more detailed comparison of jet cluster results shows that CMS results are mainly responsible for extending the range, consistent with sensitivity rankings.\\ 
\hline
3 & $t\bar{t}$ results are characterized by large $\vec{\delta}$ but there are only a few points and they are found inside the jet cluster. &
While the $t\bar{t}$ results follow similar distributions to the jet cluster, they do contain outlying points.\\
\hline
4 & Results from semi-inclusive charm production at HERA (147) are found to overlap with the DIS and jet clusters. &
These results do not take significant values in any direction of the $\vec{\delta}$ space, directional information is misleading here.\\
\hline
5 & CCFR/NuTeV dimuon SIDIS results (124-127) are orthogonal, the direction cannot be resolved in the selected view. &
The tour  resolves the orthogonal direction and further allows to identify outlying points.\\
\hline
6 & Reciprocated distance as summary statistic to characterize ``relevance'' of results. &
We can use the ranking as guidance to select results to highlight in the visualisation to gain understanding of how the summary statistics relate to raw distributions.\\
\hline
\end{tabular}}
\caption{Summary of key findings, comparing observations made with visualising PDFSense results with the TFEP and with additional insights that can be made using tour. A complete list of experimental datasets together with their CTEQ labelling IDs is given in Appendix~\ref{sec:cteqIDs}.}
\label{tab:comparison}
\end{table}

\subsection{Results from PDFSense \& TFEP}
\begin{figure}[ht!]\centering
\includegraphics[width=0.49\textwidth]{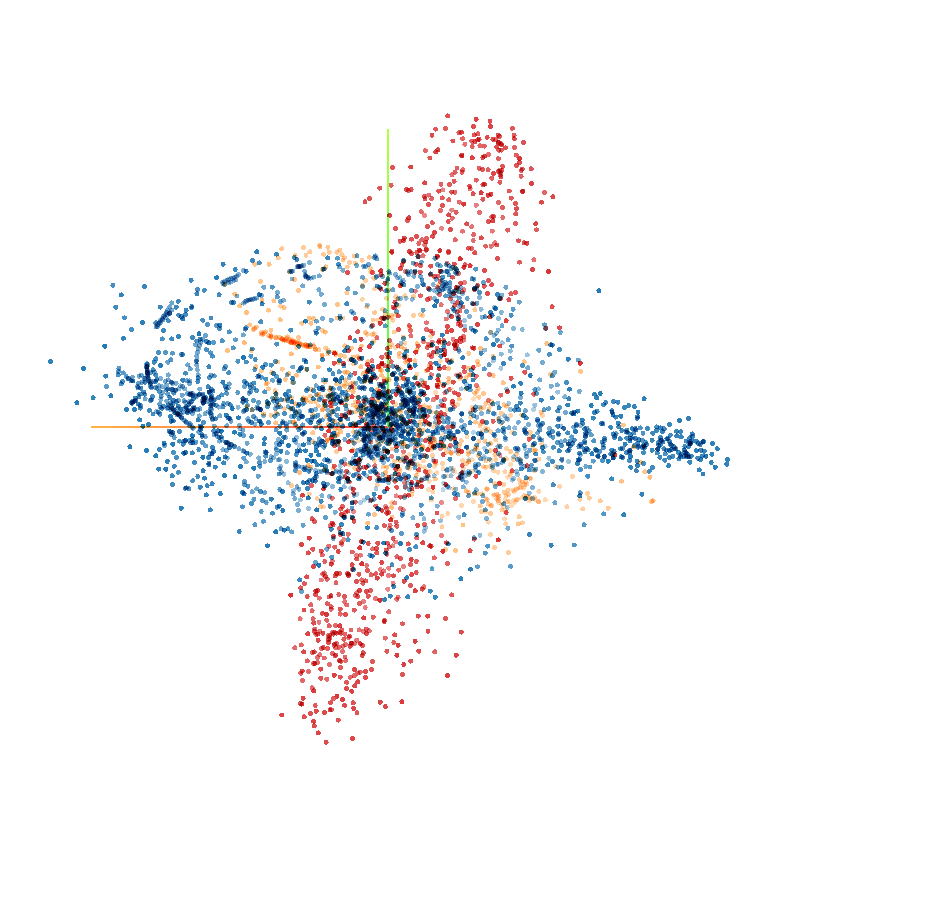}
\includegraphics[width=0.49\textwidth]{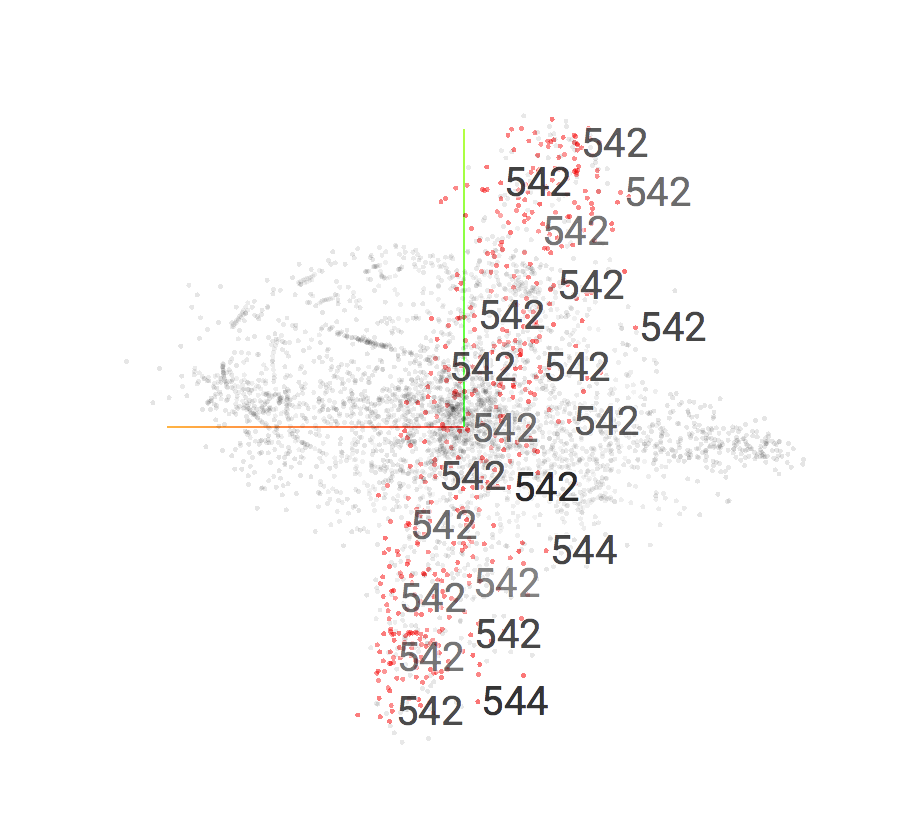}
\includegraphics[width=0.49\textwidth]{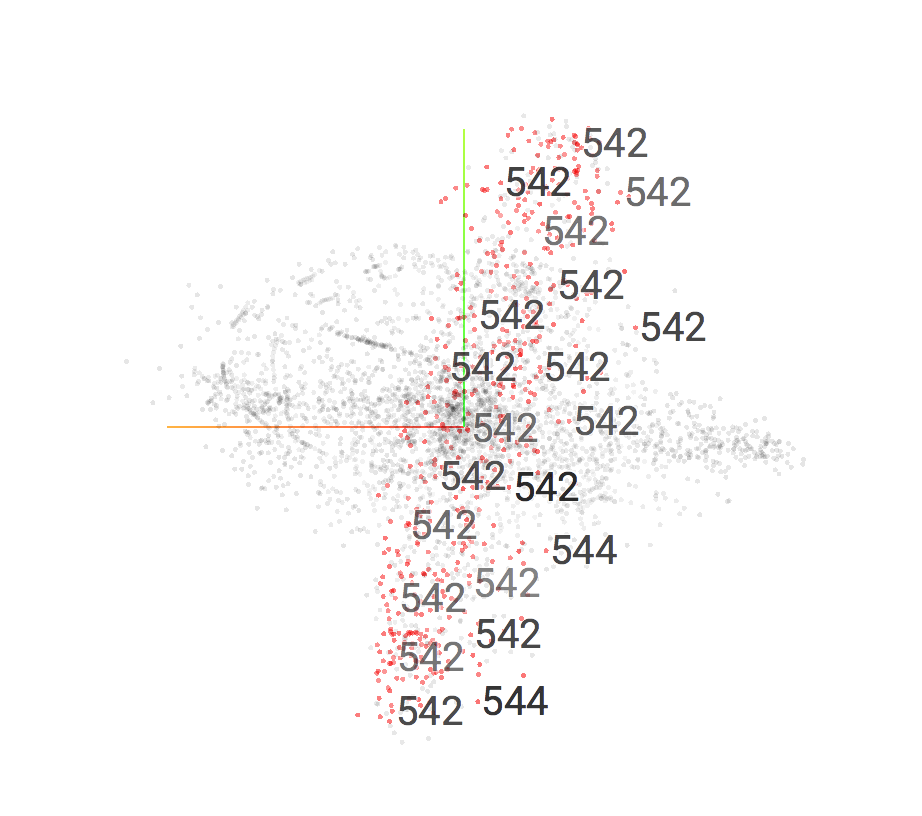}
\includegraphics[width=0.49\textwidth]{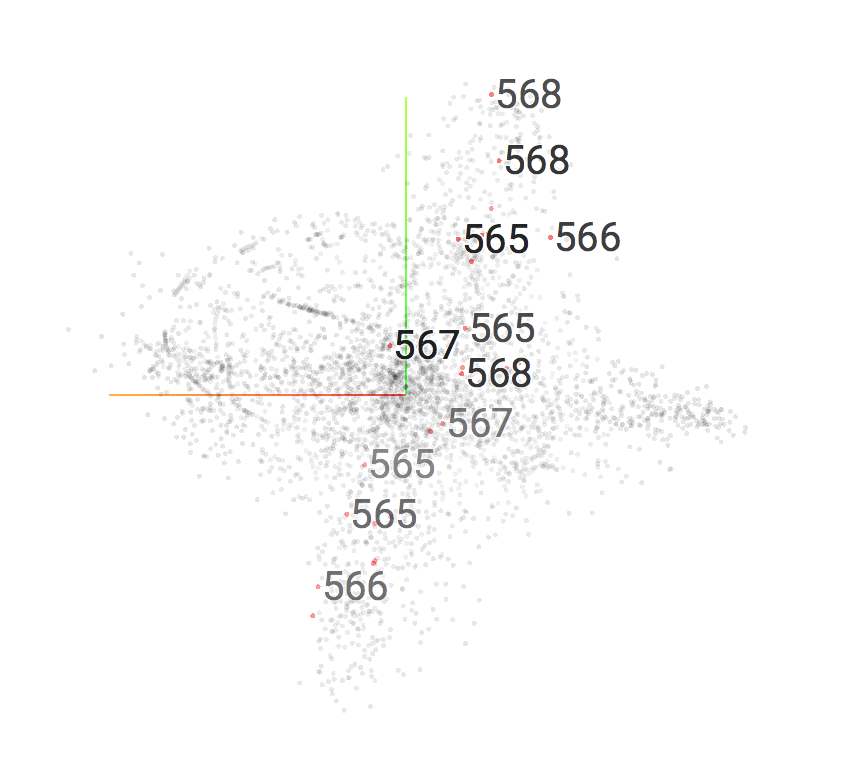}
\caption{Projections obtained with TFEP, where principal components 3, 5 and 8 have been selected, and the view was rotated such that the jet+$t\bar{t}$ cluster is roughly orthogonal to the DIS cluster. The top left plot shows grouping into jets+$t\bar{t}$ (red), DIS (blue) and VBP (orange), the remaining plots highlight subgroups (indicated by CTEQ labelling IDs shown in the appendix) of the jets+$t\bar{t}$ cluster in the same view.}
\label{fig:dataProjections}
\end{figure}

For comparison we first reproduce results similar to those found in~\cite{Wang:2018heo} by using the TFEP software. A selection of four views is shown in Figure~\ref{fig:dataProjections}, for a complete set of plots related to the PDFSense column in Table~\ref{tab:comparison} we refer the reader to~\cite{Wang:2018heo}. The selected examples show how the view was chosen based on orthogonality of assigned groups, and how for the example of the jet+$t\bar{t}$ group the various contributions have been compared.

We can identify several limitation in using the TFEP software for the visualisation:
\begin{itemize}
\item Relevant information about the distributions is encoded in more than 3 dimensions. This is clear as PCs 3, 5 and 8 have been selected in the visualisation, thus the majority of variation in the data is not captured in Figure~\ref{fig:dataProjections}. Moreover, the application of t-SNE clustering shown in~\cite{Wang:2018heo} results in a large number of clusters, indicating higher dimensional structure. It would be preferable to display it as a linear projection for which interpretations are straightforward.
\item The sphering of data points when preparing the PCA visualisation is removing relevant information about the length of the vectors $\vec{\delta}_i$.
\item In addition while the online tool allows highlighting of groups it is considerably less flexible in selecting options compared to scripted tools like the tour, limiting the detail in which the results can efficiently be studied.
\end{itemize}
We next explore how these points can be addressed, in particular in the framework of dynamical projections and the tour algorithm.

\subsection{Expanded findings made using the tour}

We first optimize the number of principal components considered in our study, and then show how the tour results expand on previous observations, as was summarized in Table~\ref{tab:comparison}. The mapping from the original $\delta$ coordinates onto the PCs for all PCAs considered in this work are listed in Appendix~\ref{sec:PCAmap}.

\subsubsection{PCA, normalisation and variance explained}

In the following we study two sets of principal components (PCA1, PCA2), corresponding to the two data preparation choices described above (i.e.\ PCA1=centered, and PCA2=centered and sphered). Results from each are compared. Note that for this problem, the centering has negligible impact on the results as the mean value in each direction $\delta_{i,l}^{\pm}$ is close to zero.

\begin{figure}[ht!]\centering
\includegraphics[width=0.5\textwidth]{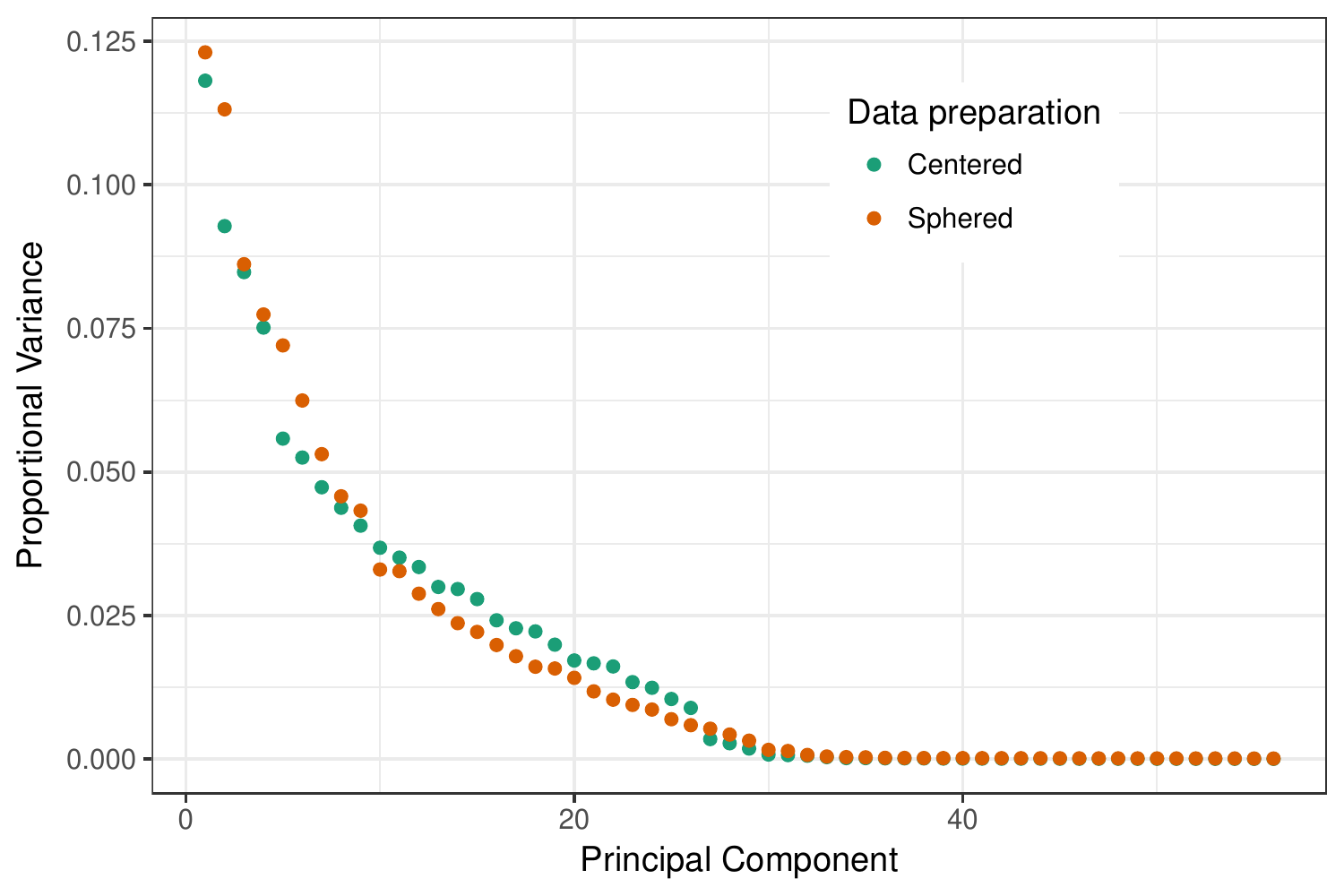}
\caption{Proportional  variance explained by the principal components of the 56 dimensional parameter space. To capture all the variation, one would need close to 30 principal components, but around 6 captures about 50\% of the variation. Both data preparations produce similar variance explanation, but the differences are enough to matter in some interpretations. 
\label{fig:variance}}
\end{figure}

An important consideration is the number of PCs that contain relevant information.
To study this we show in Figure~\ref{fig:variance} the proportional variance that is explained by the principal components,
for the two choices of the PCA, with labels ``Centered'' for PCA performed on centered data (PCA1) and
``Sphered'' for the PCA obtained for centered and sphered data (PCA2) thus reproducing results from Figure~\ref{fig:dataProjections}.
We find a steep curve for the first few PCs, followed by a slow decay of the proportional variance, and the curve only flattens out towards zero around PC30.
As a consequence we expect that looking at a 3 dimensional subset of the first 10 PCs is not sufficient to understand the variation in the considered parameter space, and that judging similarity based on the view in Figure~\ref{fig:dataProjections} only, is misleading.

In the following we want to study a higher dimensional subspace where we base the number of dimensions considered on the results found in Figure~\ref{fig:variance}. 

For simplicity, we illustrate the tour approach using just the first 6 PCs, which captures about 50\% of the overall variation.This is sufficient to provide new insights as compared to Figure~\ref{fig:dataProjections} (left), and additional PCAs can be added for detailed studies of subgroups as we do below.

\begin{table}[h]
\makebox[\textwidth]{\begin{tabular}{|c|ccccccccccccccc|}
\hline
PCA & 1 & 2 & 3 & 4 & 5 & 6 & 7 & 8 & 9 & 10 & 11 & 12 & 13 & 14 & 15\\
\hline
PCA1 & 12 & 21 & 30 & 37 & 43 & 48 & 53 & 57 & 61 & 65 & 68 & 72 & 75 & 78 & 80\\
PCA2 & 12 & 24 & 32 & 40 & 47 & 53 & 59 & 63 & 68 & 71 & 74 & 77 & 80 & 82 & 84\\
\hline
\end{tabular}}
\caption{Cumulative variance as \% explained by the first 15 PCs.}
\label{tab:cumVar}
\end{table}

\subsubsection{Grand tour result details}
A short tour path is generated, of 20 basis planes and associated interpolation between them, of 2 dimensional projections of 6-d. This is used to compare between multiple groups. 
The  examples considered are guided by  findings in~\cite{Wang:2018heo} and are summarized in Table~\ref{tab:comparison}.

\begin{figure}
\includegraphics[width=0.5\linewidth]{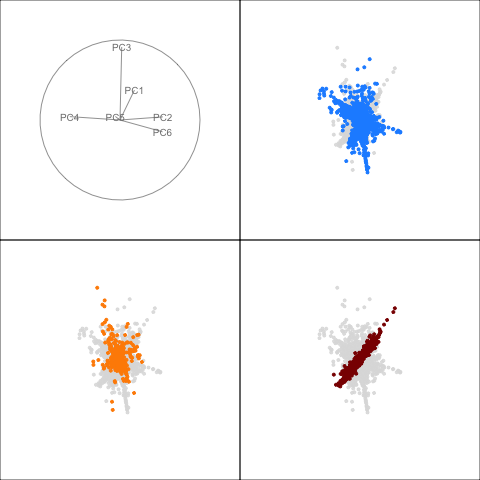}
\includegraphics[width=0.5\linewidth]{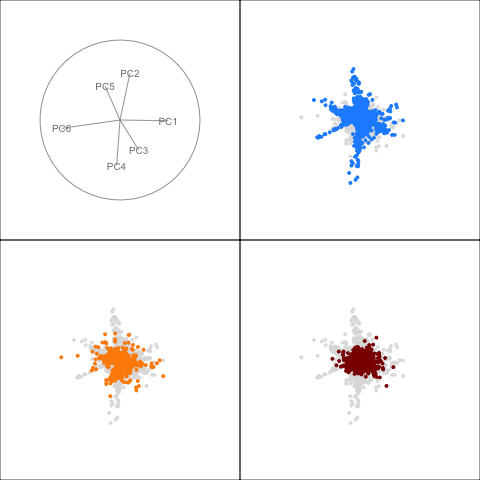}
\includegraphics[width=0.5\linewidth]{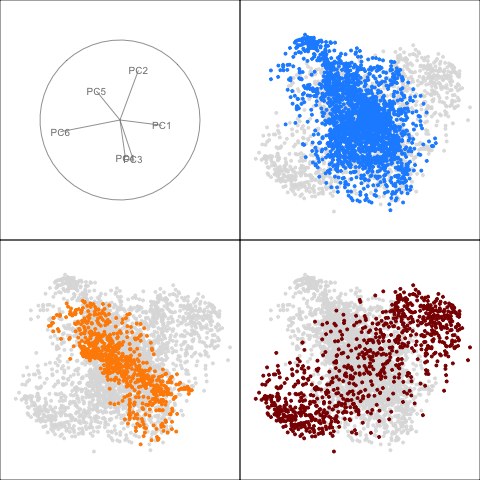}
\includegraphics[width=0.5\linewidth]{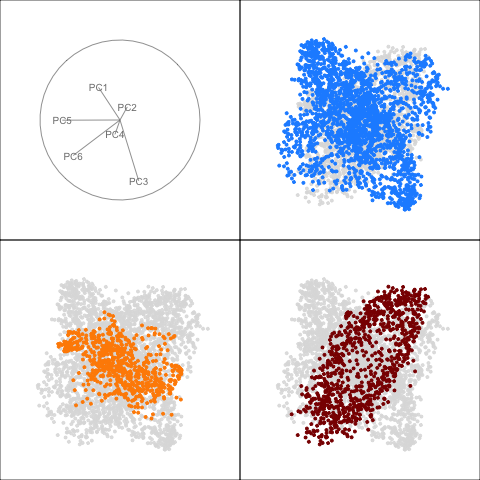}
\caption{Selected views from the grand tour results of the full dataset. The data points are grouped into DIS, VBP and jets cluster,
shown in blue, orange and red respectively. Top left plot shows the projection of the PCs, and other plots show the three subgroups. Colour indicates group, and grey shows the entire data set, as a reference in order to make comparisons between groups. PCA1  \href{http://arxiv.org/src/1806.09742v2/anc/allcenter.gif}{(see animation here)}  is shown on the top row, PCA2 \href{http://arxiv.org/src/1806.09742v2/anc/allsphere.gif}{(see animation here)} on the bottom row, the left views show a separation between DIS and jets clusters, the right views show the multidimensionality in the DIS cluster.}
\label{fig:animAll}
\end{figure}

\paragraph{Grouping of data points}
We first consider a display corresponding to Figure~\ref{fig:dataProjections} (left), i.e.\ the data set is grouped into three main clusters.
Selected views from the animation are shown in Figure~\ref{fig:animAll}, PCA1 (left) and  PCA2 (right). The same colors used in Ref. \cite{Wang:2018heo}  indicate the grouping: the DIS cluster is shown in blue, VBP in orange and the jets cluster in red.
The first window in the display shows the axes,
the other windows show the projected data, where one group is highlighted in color, while the remaining points are shown below in grey for easy comparison.
As can be seen from the selected views, in any particular static view  it is only possible to separate two of them at a  time. The static views are not sufficient to convey the full picture obtained by watching the tour animation which allows to separate all three groups. The tour indicates that there is higher dimensional structure in the data points as can be seen in the linked animation.

In addition, it is possible to visually identify substructure  within the clusters (e.g.\ groups of points aligned along some direction) as well as outlying points. 
This is especially true for PCA1 which is found to provide a much clearer picture than PCA2. 
We also find that the DIS and VBP clusters extend in multiple directions, while the jets cluster seems to be well described in a single plane.

\begin{figure}
\begin{center}
\includegraphics[width=0.7\linewidth]{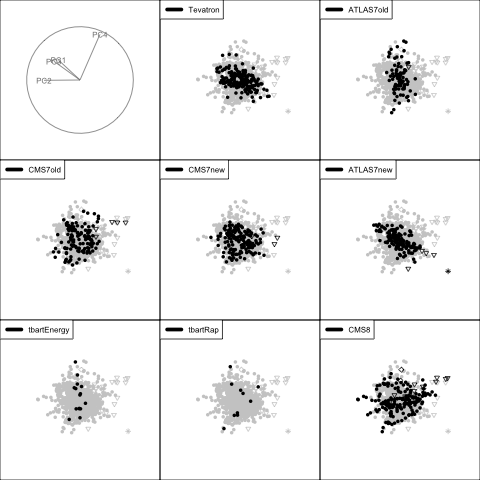}
\caption{Focusing on the jets cluster, showing only the first 4 PCs. Top left plot shows the projection coordinates, groups (Tevatron, ATLAS7old, ...) are focused in black in each plot, and grey shows all the data enabling direct comparison between subgroups. This view from the grand tour was selected because it clearly separates the outlying point in the ATLAS7new dataset. In addition the view also illustrates how the CMS results extend the reach away from the main cluster \href{http://arxiv.org/src/1806.09742v2/anc/jetCluster.gif}{(see animation here)} .}
\label{fig:jetCluster}
\end{center}
\end{figure}

\paragraph{The jet cluster}
In more detail, we investigate the jet cluster. These results are of special interest since they contain indeed the largest data sets to be added in the fit, which were indeed found to be important according to~\cite{Wang:2018heo}.
In addition, the new data from LHC jet measurements is of interest because of possible tensions  e.g. \cite{talk1,talk2}.
As seen above the jet cluster appears to be described in a lower dimensional subspace. Indeed performing PCA on the results in the jet cluster alone we see that the cumulative proportional variance reaches 49/75/91/95 \% for PC1/2/3/4 respectively, with the proportional variance dropping to less than 2\% for PC5. We therefore study substructure in this 4 dimensional space. While~\cite{Wang:2018heo} distinguish three types of groups, i.e.\ ``old'' jet results (those included in the CT14HERA2 fit), ``new'' jet results (more recent ATLAS and CMS results) and $t\bar{t}$, it makes sense to differentiate the LHC results further by experiment and $\sqrt{s}$ (motivated also by the differences in sensitivities observed in~\cite{Wang:2018heo}). For simplicity we consider only the results from performing PCA on the centered data shown in Figure~\ref{fig:jetCluster} with grouping into: Tevatron (IDs 504, 515), ATLAS7old (535), CMS7old (538), CMS7new (542), ATLAS7new (544), $t\bar{t}$-energy (565, 567), $t\bar{t}$-rap (566, 568) and CMS8 (545). Indeed we observe that the Tevatron results as well as the ATLAS results generally fall in the center of the cluster, with exception of some outlying points. On the other hand CMS  7 and 8 TeV results extend in (different) new directions. It is interesting to note that ``old'' CMS 7 TeV results extend further out than the corresponding ``new'' ones. In fact while the new measurement extended to higher rapidities and lower values in jet $p_T$, the old measurement contains higher $p_T$ bins at no longer present in the updated result, which turn out to give large values of $\vec{\delta}$.
Finally for $t\bar{t}$ results we distinguish the observations binned in energy ($p_T^t$ or $m_{t\bar{t}}$) or rapidity ($y_{\langle t/\bar{t}\rangle}$ or $y_{t\bar{t}}$). We can identify differences between the two groups in the visualisation, however as already noted in~\cite{Wang:2018heo} the data points are not significantly different from the main jet cluster.

It is interesting to study which data points are found to be outlying in the visualisation. These points are highlighted in Figure~\ref{fig:jetCluster} and are best distinguished when watching the tour animation:
\begin{itemize}
\item $|y| > 2.5$ and $\mu > 950$~GeV -- marked with a star symbol: only one such point is found in the 7 TeV data sets. It occurs in ATLAS7new, it is the last rapidity bin and is clearly outlying  (large negative values in PCs 1, 2 and 3). However no particular trend is observed when comparing with points in nearby bins. There are two more such data points in the CMS8 data set, but they do not stand out in $\delta$ space. 
\item $|y| > 2$ and $\mu > 1000$~GeV -- marked with downward pointing triangle. These points are seen to align in a new direction, away from the main cluster highlighting their importance in the fits.

They are also useful for comparing the different CMS results: in this case there are common points to both datasets that nevertheless look different, suggesting the need for further study of these points.

\item for CMS8 we also highlight $|y| < 1$ and $\mu < 200$ -- marked with diamond symbol: they are very different from the main distribution and give large positive values in PC1. It is interesting that we can clearly separate these low $\mu$ bins in CMS8 set but not in CMS7.
\end{itemize}

\paragraph{The DIS cluster}
We next consider subgroups of the DIS cluster for which the TFEP visualisation allowed only limited interpretation. Concretely, while the bulk of the cluster was clearly spanned by the HERA results (ID 160) as expected, other results were found to follow quite different distributions. In particular the Charm SIDIS (ID 147) results are distributed in a different direction, overlapping partly with both the DIS and the jet clusters, while the dimuon SIDIS results (IDs 124-127) were found in the center of the distribution and it was concluded that this cluster extends in an orthogonal direction, although it was not shown explicitly.

\begin{figure}
\includegraphics[width=0.5\linewidth]{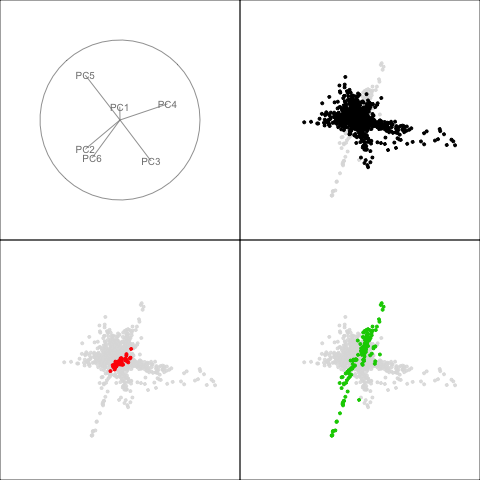}
\includegraphics[width=0.5\linewidth]{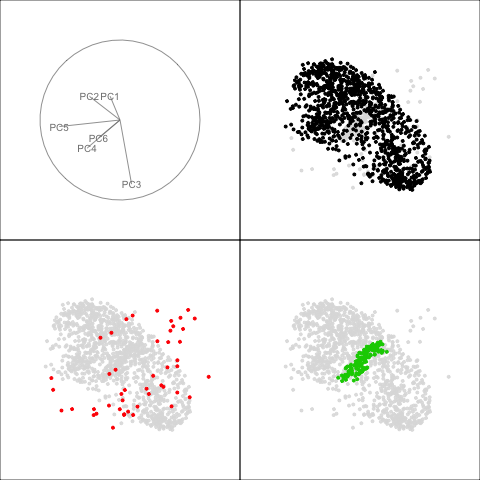}
\caption{As Figure~\ref{fig:animAll}, but showing only selected results in the DIS cluster, i.e.\ DIS HERA1+2 (black), Charm SIDIS (red) and dimuon SIDIS (green). The left view is for PCA1 \href{http://arxiv.org/src/1806.09742v2/anc/DIScenter.gif}{(see animation here)}  shows clear separation of dimuon SIDIS results, the right view for PCA2 \href{http://arxiv.org/src/1806.09742v2/anc/DISsphere.gif}{(see animation here)} shows apparent separation of charm SIDIS results obtained by focussing on directional information.}
\label{fig:animDIS}
\end{figure}

We therefore compare in detail these three groups. In this case it is useful to consider both  PCA1 and PCA2, the latter more closely related to the TFEP output.
First, we observe that the dimuon SIDIS is poorly separated in the PCA2 projection, whereas PCA1 clearly shows how it extends considerably away from the main DIS cluster (ID 160). 
On the other hand, the charm SIDIS can be separated more easily when studying the directional information in the PCA2 projection  because the individual values in the space of deltas are all comparatively small. 
These results suggest that either predictions for these type of observables are well under control in the existing fits, or that alternatively the experimental errors are too large for them to be constraining.
We also observe substructure in the DIS HERA1+2, see Figure~\ref{fig:animDIS} and the corresponding animation, indicating that this group combines a number of qualitatively different types of results.

\begin{figure}
\begin{center}
\includegraphics[width=0.49\linewidth]{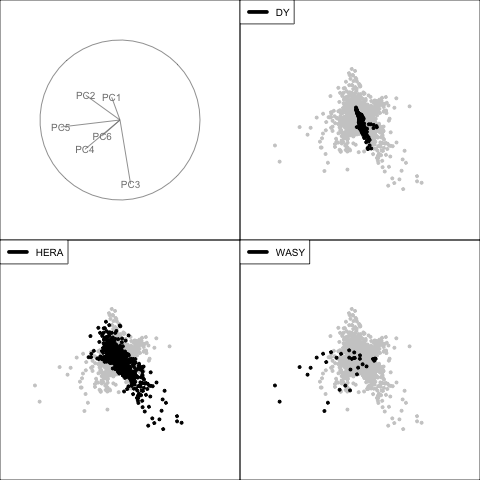}
\includegraphics[width=0.49\linewidth]{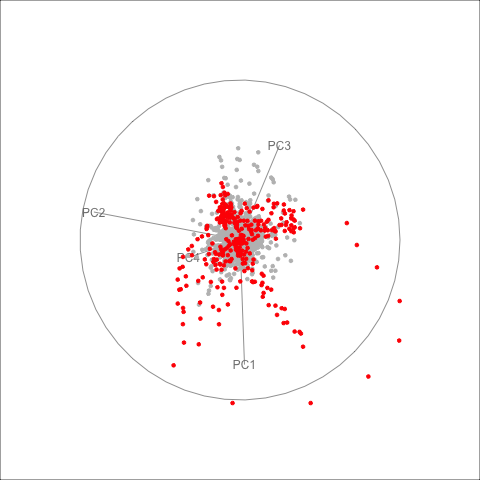}
\caption{Left: Comparison of groups with large reciprocated distance measures, where now the full dataset is shown below in gray. Right: Comparison in subspace found by performing PCA on DY data only, where DY data is shown in red and all other data points are shown below in gray. Again selected views from the grand tour results are shown here. The left view \href{http://arxiv.org/src/1806.09742v2/anc/grDYcenter.gif}{(see animation here)} roughly shows how the HERA and WASY data points are far away from the main distribution of data points, while the DY points are found only in the center. The right view \href{http://arxiv.org/src/1806.09742v2/anc/DYvsAll.gif}{(see animation here)} illustrates the three different types of distributions found in the DY group.}
\label{fig:animRecDist}
\end{center}
\end{figure}

\paragraph{Comparison with summary statistics}
We now consider the experimental results with the highest values in reciprocated distances to show they can also be easily distinguished with our
visualisation.
We  highlight three groups in Figure~\ref{fig:animRecDist}: the HERA dataset (ID 160), the W asymmetry measurements (ID 234, 266 and 281) and the fixed-target Drell-Yan measurements from E605 and E866 (ID 201, 203 and 204).

Indeed we find that the W asymmetry measurements (234, 266 and 281) follow a very distinct distribution, as does the HERA DIS dataset (160). On the other hand, the fixed-target Drell-Yan measurements (201, 203 and 204),  do not stand out  in our visualisation. We find that this is a consequence of the dimension reduction,\footnote{Recall that the selected first six PCs only capture 48\% of overall variance} and we can easily identify views separating this group from the other data points when considering additional dimensions. Here we show this by looking at projections found by performing PCA on this data subset only and using it to compare it to the other data sets in the subspace of the first 4 PCs thus defined. Note however that the tour allows visualisation of the distributions in the full parameter space which would yield the same information. Our choice of procedure is simply to limit the viewing times required, which grow with the number of dimensions considered.\footnote{When working in the full parameter space one should consider the definition of projection pursuit indices to guide the tour to interesting views, one may e.g.\ define an index that finds views where a selected group of data points is maximally separated from the cluster of points, similar to the definition of reciprocated distances.}

This type of visualisation, together with inverting the mapping onto principal components, may be used to identify the origin (i.e.\ underlying physics) of the large differences. 
For example the first three PCs found for the DY dataset capture three different distributions, and mapping those back to the original $\delta$ directions together with study of those directions with respect to uncertainty in individual parton pdfs may provide additional insight. Such detailed investigations are however beyond the scope of this study.

\section{Summary and conclusions}
\label{sec:Conclusions}

Starting from the set of 56 dimensional vectors in the space of residual responses calculated in ~\cite{Wang:2018heo}, we have demonstrated how the grand tour may be used for visualizations in particle physics.  
The 56 dimensions are reduced to 6 dimensions (for illustration) using principal component analysis, and the resulting representation is then passed onto the tour.  The findings made about the fits using the tour, even with only 6 dimensions, are more comprehensive and clearer than what TFEP allows.

The tour visualisation verified several results from~\cite{Wang:2018heo}, notably, the separation between DIS, VBP and JET experiments into clusters populating different regions of delta space. It also allowed us to go into further detail by examining certain substructures within these groups. We have moreover demonstrated that the tour can complement and support analyses based on the use of reciprocated distances.

In our examples we have considered performing the PCA either on centered data (PCA1) or on centered and sphered data (PCA2), as they highlight different aspects of the structure, the former retaining length information and the latter emphasizing directionality. 
In general we find the results from PCA1 more useful, in particular for this application where the length of the individual data point vectors (i.e.\ for each experiment)
carries important information that is lost when sphering the input data.

The sensitivity defined in ~\cite{Wang:2018heo}, or projection of $\delta$s onto a direction given by the gradient of a QCD variable (e.g. cross section prediction) can also be inspected visually and the tour permits this visualisation in multiple dimensions. 

We conclude that the above described method is a valuable tool for PDF uncertainty and sensitivity studies. In addition,  the visual analysis allows a better understanding of the method itself and can uncover unexpected features, and even  possibly errors. It can provide experiments with a guide to the measurements needed to improve PDF fits.

\section*{Acknowledgements}

This work was supported in part by the Australian Research Council. We thank  Nicholas Spyrison  for help with the animations and Timothy Hobbs and Fred Olness for clarifications on their work.

\appendix

\section{CTEQ labelling IDs}
\label{sec:cteqIDs}

These are the same numbers used in Ref.~\cite{Wang:2018heo}, we reproduce them here for convenience. Experimental datasets included in the CT14HERA2 fit are listed in Table~\ref{tab:EXP_1}, additional results included in the study are given in Table~\ref{tab:EXP_3}.

\begin{table}
\begin{tabular}{|l|lr|c|}
\hline
\textbf{ID\#}  & \textbf{Experimental dataset}  &  & \textbf{Group} \tabularnewline
\hline
101  & BCDMS $F_{2}^{p}$  & \cite{Benvenuti:1989rh}  & DIS \tabularnewline
\hline
102  & BCDMS $F_{2}^{d}$  & \cite{Benvenuti:1989fm}  & DIS \tabularnewline
\hline
104  & NMC $F_{2}^{d}/F_{2}^{p}$  & \cite{Arneodo:1996qe}  & DIS \tabularnewline
\hline
108  & CDHSW $F_{2}^{p}$  & \cite{Berge:1989hr}  & DIS \tabularnewline
\hline
109  & CDHSW $F_{3}^{p}$  & \cite{Berge:1989hr}  & DIS \tabularnewline
\hline
110  & CCFR $F_{2}^{p}$  & \cite{Yang:2000ju}  & DIS \tabularnewline
\hline
111  & CCFR $xF_{3}^{p}$  & \cite{Seligman:1997mc}  & DIS \tabularnewline
\hline
124  & NuTeV $\nu\mu\mu$ SIDIS  & \cite{Mason:2006qa}  & DIS \tabularnewline
\hline
125  & NuTeV $\bar{\nu}\mu\mu$ SIDIS  & \cite{Mason:2006qa}  & DIS \tabularnewline
\hline
126  & CCFR $\nu\mu\mu$ SIDIS  & \cite{Goncharov:2001qe}  & DIS \tabularnewline
\hline
127  & CCFR $\bar{\nu}\mu\mu$ SIDIS  & \cite{Goncharov:2001qe}  & DIS \tabularnewline
\hline
145  & H1 $\sigma_{r}^{b}$ ($57.4\mbox{ pb}^{-1}$)  & \cite{Aktas:2004az}\cite{Aktas:2005iw}  & DIS \tabularnewline
\hline
147  & Combined HERA charm production ($1.504\mbox{ fb}^{-1}$)  & \cite{Abramowicz:1900rp}  & DIS \tabularnewline
\hline
160  & HERA1+2 Combined NC and CC DIS ($1\mbox{ fb}^{-1}$)  & \cite{Abramowicz:2015mha}  & DIS \tabularnewline
\hline
169  & H1 $F_{L}$ ($121.6\mbox{ pb}^{-1}$)  & \cite{Collaboration:2010ry}  & DIS \tabularnewline
\hline
201  & E605 DY  & \cite{Moreno:1990sf}  & VBP \tabularnewline
\hline
203  & E866 DY, $\sigma_{pd}/(2\sigma_{pp})$  & \cite{Towell:2001nh}  & VBP \tabularnewline
\hline
204  & E866 DY, $Q^{3}d^{2}\sigma_{pp}/(dQdx_{F})$  & \cite{Webb:2003ps}  & VBP \tabularnewline
\hline
225  & CDF Run-1 $A_{e}(\eta^{e})$ ($110\mbox{ pb}^{-1}$)  & \cite{Abe:1996us}  & VBP \tabularnewline
\hline
227  & CDF Run-2 $A_{e}(\eta^{e})$ ($170\mbox{ pb}^{-1}$)  & \cite{Acosta:2005ud}  & VBP \tabularnewline
\hline
234  & D$\emptyset$~ Run-2 $A_{\mu}(\eta^{\mu})$ ($0.3\mbox{ fb}^{-1}$)  & \cite{Abazov:2007pm}  & VBP \tabularnewline
\hline
240  & LHCb 7 TeV $W/Z$ muon forward-$\eta$ Xsec ($35\mbox{ pb}^{-1}$)  & \cite{Aaij:2012vn}  & VBP \tabularnewline
\hline
241  & LHCb 7 TeV $W$ $A_{\mu}(\eta^{\mu})$ ($35\mbox{ pb}^{-1}$)  & \cite{Aaij:2012vn}  & VBP \tabularnewline
\hline
260  & D$\emptyset$~ Run-2 $Z$ $d\sigma/dy_{Z}$ ($0.4\mbox{ fb}^{-1}$)  & \cite{Abazov:2006gs}  & VBP \tabularnewline
\hline
261  & CDF Run-2 $Z$ $d\sigma/dy_{Z}$ ($2.1\mbox{ fb}^{-1}$)  & \cite{Aaltonen:2010zza}  & VBP \tabularnewline
\hline
266  & CMS 7 TeV $A_{\mu}(\eta)$ ($4.7\mbox{ fb}^{-1}$)  & \cite{Chatrchyan:2013mza}  & VBP \tabularnewline
\hline
267  & CMS 7 TeV $A_{e}(\eta)$ ($0.840\mbox{ fb}^{-1}$)  & \cite{Chatrchyan:2012xt}  & VBP \tabularnewline
\hline
268  & ATLAS 7 TeV $W/Z$ Xsec, $A_{\mu}(\eta)$ ($35\mbox{ pb}^{-1}$)  & \cite{Aad:2011dm}  & VBP \tabularnewline
\hline
281  & D$\emptyset$~ Run-2 $A_{e}(\eta)$ ($9.7\mbox{ fb}^{-1}$)  & \cite{D0:2014kma}  & VBP \tabularnewline
\hline
504  & CDF Run-2 incl. jet ($d^2\sigma/dp_{T}^{j}dy_{j}$) ($1.13\mbox{ fb}^{-1}$)  & \cite{Aaltonen:2008eq}  & JET \tabularnewline
\hline
514  & D$\emptyset$~ Run-2 incl. jet ($d^2\sigma/dp_{T}^{j}dy_{j}$) ($0.7\mbox{ fb}^{-1}$)  & \cite{Abazov:2008ae}  & JET \tabularnewline
\hline
535  & ATLAS 7 TeV incl. jet ($d^2\sigma/dp_{T}^{j}dy_{j}$) ($35\mbox{ pb}^{-1}$)  & \cite{Aad:2011fc}  & JET \tabularnewline
\hline
538  & CMS 7 TeV incl. jet ($d^2\sigma/dp_{T}^{j}dy_{j}$) ($5\mbox{ fb}^{-1}$)  & \cite{Chatrchyan:2012bja}  & JET \tabularnewline
\hline
\end{tabular}
\caption{Experimental datasets considered as part of CT14HERA2 and included in the analysis. IDs are following the standard CTEQ labelling system with 1XX/2XX/5XX representing datasets in the DIS/VBP/JET group.\label{tab:EXP_1}}
\end{table}

\begin{table}
\begin{tabular}{|l|lr|c|}
\hline
\textbf{ID\#}  & \textbf{Experimental dataset}  &  & \textbf{Group} \tabularnewline
\hline
\hline
        \textbf{245}  & LHCb 7 TeV Z/W muon forward-$\eta$ Xsec ($1.0\mbox{ fb}^{-1}$)  & \cite{Aaij:2015gna}  & VBP \tabularnewline
\hline
        \textbf{246}  & LHCb 8 TeV Z electron forward-$\eta$ $d\sigma/dy_{Z}$ ($2.0\mbox{ fb}^{-1}$)  & \cite{Aaij:2015vua}  & VBP \tabularnewline
\hline
        \textbf{247}  & ATLAS 7 TeV $d\sigma/dp_{T}^{Z}$ ($4.7\mbox{ fb}^{-1}$)  & \cite{Aad:2014xaa}  & VBP \tabularnewline
\hline
        \textbf{249}  & CMS 8 TeV W muon, Xsec, $A_{\mu}(\eta^{\mu})$ ($18.8\mbox{ fb}^{-1}$)  & \cite{Khachatryan:2016pev}  & VBP \tabularnewline
\hline
        \textbf{250}  & LHCb 8 TeV W/Z muon, Xsec, $A_{\mu}(\eta^{\mu})$ ($2.0\mbox{ fb}^{-1}$)  & \cite{Aaij:2015zlq}  & VBP \tabularnewline
\hline
        \textbf{252}  & ATLAS 8 TeV Z ($d^{2}\sigma/d|y|_{ll}dm_{ll}$) ($20.3\mbox{ fb}^{-1}$)  & \cite{Aad:2016zzw}  & VBP \tabularnewline
\hline
        \textbf{253}  & ATLAS 8 TeV ($d^{2}\sigma/dp_{T}^{Z}dm_{ll}$) ($20.3\mbox{ fb}^{-1}$)  & \cite{Aad:2015auj}  & VBP \tabularnewline
\hline
        \textbf{542}  & CMS 7 TeV incl. jet, R=0.7, ($d^2\sigma/dp_{T}^{j}dy_{j}$) ($5\mbox{ fb}^{-1}$)  & \cite{Chatrchyan:2014gia}  & JET \tabularnewline
\hline
        \textbf{544}  & ATLAS 7 TeV incl. jet, R=0.6, ($d^2\sigma/dp_{T}^{j}dy_{j}$) ($4.5\mbox{ fb}^{-1}$)  & \cite{Aad:2014vwa}  & JET \tabularnewline
\hline
        \textbf{545}  & CMS 8 TeV incl. jet, R=0.7, ($d^2\sigma/dp_{T}^{j}dy_{j}$) ($19.7\mbox{ fb}^{-1}$)  & \cite{Khachatryan:2016mlc}  & JET \tabularnewline
\hline
        \textbf{565}  & ATLAS 8 TeV $t\overline{t}\:d\sigma/dp_{T}^{t}$ ($20.3\mbox{ fb}^{-1}$)  & \cite{Aad:2015mbv}  & JET \tabularnewline
\hline
        \textbf{566}  & ATLAS 8 TeV $t\overline{t}\:d\sigma/dy_{<t/\overline{t}>}$ ($20.3\mbox{ fb}^{-1}$)  & \cite{Aad:2015mbv}  & JET \tabularnewline
\hline
        \textbf{567}  & ATLAS 8 TeV $t\overline{t}\:d\sigma/dm_{t\overline{t}}$ ($20.3\mbox{ fb}^{-1}$)  & \cite{Aad:2015mbv}  & JET \tabularnewline
\hline
        \textbf{568}  & ATLAS 8 TeV $t\overline{t}\:d\sigma/dy_{t\overline{t}}$ ($20.3\mbox{ fb}^{-1}$)  & \cite{Aad:2015mbv}  & JET \tabularnewline
\hline
\end{tabular}\caption{Same as Table~\ref{tab:EXP_1}, but showing experimental datasets not incorporated in the CT14HERA2 fit but included
in the augmented CTEQ-TEA set. \label{tab:EXP_3} }
\end{table}

\section{Projections on the principal components}
\label{sec:PCAmap}
Here we list the projection matrices used to obtain projections in the various PC subspaces considered, for PCA1 in Table~\ref{tab:PCA1}, for PCA2 in Table~\ref{tab:PCA2}, for the PCA performed on the jet cluster in Table~\ref{tab:PCAjet} and for the PCA performed only on the DY data in Table~\ref{tab:PCAdy}.

\begin{table}[t]
\begin{tabular}{|r|rrrrrr|}
  \hline
 & PC1 & PC2 & PC3 & PC4 & PC5 & PC6 \\ 
  \hline
X1 & 0.14 & -0.22 & 0.12 & -0.15 & 0.07 & -0.09 \\ 
  X2 & 0.11 & -0.33 & 0.10 & -0.21 & 0.14 & -0.02 \\ 
  X3 & -0.10 & -0.14 & 0.10 & 0.12 & -0.25 & -0.15 \\ 
  X4 & -0.22 & -0.15 & -0.02 & -0.03 & 0.16 & 0.11 \\ 
  X5 & -0.17 & -0.08 & 0.15 & 0.11 & -0.12 & -0.08 \\ 
  X6 & -0.18 & -0.09 & 0.15 & 0.05 & -0.14 & -0.07 \\ 
  X7 & 0.03 & 0.34 & 0.33 & -0.10 & -0.18 & -0.14 \\ 
  X8 & -0.25 & -0.00 & 0.15 & 0.14 & 0.11 & 0.11 \\ 
  X9 & -0.13 & -0.02 & 0.11 & -0.20 & -0.25 & 0.07 \\ 
  X10 & 0.03 & -0.06 & -0.01 & -0.03 & 0.05 & -0.08 \\ 
  X11 & -0.03 & -0.09 & 0.00 & -0.16 & -0.04 & -0.10 \\ 
  X12 & 0.02 & -0.01 & 0.05 & 0.04 & -0.00 & 0.05 \\ 
  X13 & -0.16 & -0.08 & 0.07 & 0.05 & -0.24 & 0.09 \\ 
  X14 & -0.05 & -0.13 & 0.11 & -0.01 & 0.20 & -0.23 \\ 
  X15 & 0.01 & -0.01 & -0.08 & 0.21 & 0.23 & -0.30 \\ 
  X16 & -0.04 & -0.04 & 0.10 & -0.20 & -0.21 & 0.28 \\ 
  X17 & -0.02 & -0.10 & -0.05 & -0.17 & -0.09 & 0.02 \\ 
  X18 & 0.00 & 0.11 & 0.05 & 0.15 & 0.08 & -0.03 \\ 
  X19 & 0.12 & -0.03 & 0.23 & 0.25 & 0.05 & 0.21 \\ 
  X20 & -0.15 & 0.03 & -0.22 & -0.27 & -0.03 & -0.17 \\ 
  X21 & -0.03 & 0.32 & -0.09 & -0.20 & 0.12 & 0.03 \\ 
  X22 & -0.03 & -0.29 & 0.12 & 0.15 & -0.12 & -0.04 \\ 
  X23 & 0.23 & -0.17 & -0.06 & 0.08 & 0.00 & 0.16 \\ 
  X24 & -0.23 & 0.22 & 0.15 & -0.07 & -0.04 & -0.15 \\ 
  X25 & -0.06 & -0.03 & -0.31 & 0.08 & -0.30 & -0.14 \\ 
  X26 & -0.02 & 0.07 & 0.37 & -0.08 & 0.32 & 0.16 \\ 
  X27 & 0.10 & 0.07 & -0.03 & -0.01 & -0.01 & 0.26 \\ 
  X28 & -0.11 & -0.08 & 0.02 & 0.01 & 0.03 & -0.26 \\ 
\hline
\end{tabular}
\hfill
\begin{tabular}{|r|rrrrrr|}
  \hline
 & PC1 & PC2 & PC3 & PC4 & PC5 & PC6 \\
  \hline
  X29 & 0.01 & 0.03 & -0.04 & 0.14 & 0.11 & -0.11 \\ 
  X30 & -0.01 & -0.01 & 0.02 & -0.06 & -0.06 & 0.05 \\ 
  X31 & 0.02 & -0.00 & 0.05 & -0.02 & 0.05 & 0.06 \\ 
  X32 & -0.03 & 0.00 & -0.05 & 0.02 & -0.05 & -0.06 \\ 
  X33 & -0.00 & -0.09 & -0.02 & -0.05 & 0.04 & 0.08 \\ 
  X34 & -0.04 & 0.08 & 0.04 & 0.06 & -0.03 & -0.08 \\ 
  X35 & -0.02 & -0.01 & -0.01 & 0.01 & 0.03 & -0.09 \\ 
  X36 & 0.02 & 0.00 & 0.00 & -0.01 & -0.02 & 0.14 \\ 
  X37 & -0.00 & -0.03 & 0.18 & -0.06 & 0.03 & 0.09 \\ 
  X38 & 0.01 & 0.04 & -0.10 & 0.05 & -0.03 & -0.07 \\ 
  X39 & -0.06 & 0.05 & -0.01 & -0.03 & 0.06 & 0.00 \\ 
  X40 & -0.14 & 0.05 & -0.03 & 0.03 & -0.13 & 0.38 \\ 
  X41 & -0.11 & 0.29 & 0.24 & -0.03 & 0.03 & 0.04 \\ 
  X42 & 0.07 & -0.09 & 0.02 & 0.01 & -0.13 & -0.09 \\ 
  X43 & -0.08 & 0.02 & -0.04 & 0.18 & -0.12 & 0.10 \\ 
  X44 & -0.05 & 0.06 & -0.04 & -0.19 & 0.19 & 0.00 \\ 
  X45 & -0.09 & 0.02 & 0.09 & -0.20 & 0.11 & -0.07 \\ 
  X46 & -0.04 & 0.18 & -0.10 & 0.40 & -0.03 & 0.06 \\ 
  X47 & -0.31 & -0.12 & 0.02 & -0.00 & -0.04 & 0.03 \\ 
  X48 & -0.20 & -0.11 & 0.08 & 0.01 & 0.13 & 0.05 \\ 
  X49 & -0.24 & 0.02 & -0.13 & -0.12 & -0.04 & 0.06 \\ 
  X50 & -0.24 & -0.00 & -0.19 & -0.21 & -0.09 & 0.04 \\ 
  X51 & -0.25 & -0.13 & -0.10 & 0.14 & 0.15 & 0.21 \\ 
  X52 & -0.13 & -0.26 & -0.17 & 0.02 & 0.21 & 0.15 \\ 
  X53 & -0.19 & 0.20 & -0.15 & 0.18 & 0.13 & 0.03 \\ 
  X54 & -0.21 & 0.07 & -0.19 & -0.07 & 0.27 & 0.00 \\ 
  X55 & -0.19 & -0.07 & 0.17 & 0.09 & 0.00 & -0.10 \\ 
  X56 & 0.23 & 0.11 & -0.20 & -0.10 & -0.01 & 0.14 \\ 
   \hline
\end{tabular}
\caption{Rotation matrix for projection on the first 6 PCs of PCA1 where X1 corresponds to eigenvector set 1 and so on.}
\label{tab:PCA1}
\end{table}

\begin{table}[t]
\begin{tabular}{|r|rrrrrr|}
  \hline
 & PC1 & PC2 & PC3 & PC4 & PC5 & PC6 \\ 
  \hline
X1 & 0.23 & -0.05 & 0.27 & 0.06 & -0.09 & 0.09 \\ 
  X2 & 0.20 & 0.05 & 0.25 & 0.13 & -0.07 & 0.14 \\ 
  X3 & -0.02 & 0.09 & 0.22 & -0.15 & 0.10 & -0.22 \\ 
  X4 & -0.14 & 0.17 & -0.01 & -0.02 & -0.25 & 0.34 \\ 
  X5 & -0.09 & -0.01 & 0.22 & -0.12 & 0.03 & -0.20 \\ 
  X6 & -0.15 & 0.05 & 0.15 & -0.07 & -0.04 & 0.01 \\ 
  X7 & -0.15 & -0.48 & -0.00 & -0.02 & -0.10 & -0.16 \\ 
  X8 & -0.30 & 0.07 & 0.24 & -0.02 & 0.16 & -0.00 \\ 
  X9 & -0.06 & -0.14 & 0.17 & -0.06 & -0.06 & -0.01 \\ 
  X10 & 0.04 & 0.03 & 0.03 & 0.04 & 0.00 & -0.05 \\ 
  X11 & 0.00 & 0.00 & 0.05 & 0.03 & -0.09 & -0.06 \\ 
  X12 & 0.03 & -0.02 & 0.06 & -0.02 & 0.07 & 0.04 \\ 
  X13 & -0.03 & 0.06 & 0.10 & -0.01 & -0.08 & -0.13 \\ 
  X14 & -0.07 & 0.02 & 0.12 & -0.09 & 0.04 & 0.16 \\ 
  X15 & 0.02 & 0.09 & -0.04 & 0.08 & -0.07 & -0.16 \\ 
  X16 & -0.03 & -0.07 & 0.05 & -0.09 & 0.06 & 0.21 \\ 
  X17 & 0.04 & -0.02 & 0.06 & -0.06 & 0.04 & -0.04 \\ 
  X18 & -0.06 & 0.01 & -0.05 & 0.05 & -0.04 & 0.03 \\ 
  X19 & -0.00 & -0.10 & 0.07 & -0.34 & 0.34 & 0.20 \\ 
  X20 & -0.03 & 0.10 & -0.05 & 0.35 & -0.30 & -0.18 \\ 
  X21 & -0.16 & -0.14 & -0.18 & 0.04 & 0.14 & 0.16 \\ 
  X22 & 0.09 & 0.11 & 0.19 & -0.07 & -0.11 & -0.12 \\ 
  X23 & 0.22 & 0.11 & 0.13 & 0.06 & -0.05 & -0.05 \\ 
  X24 & -0.28 & -0.20 & -0.07 & -0.07 & 0.05 & -0.01 \\ 
  X25 & 0.02 & 0.17 & -0.19 & -0.39 & -0.04 & -0.16 \\ 
  X26 & -0.12 & -0.20 & 0.21 & 0.38 & 0.05 & 0.18 \\ 
  X27 & 0.05 & -0.01 & -0.07 & 0.18 & 0.18 & -0.06 \\ 
  X28 & -0.07 & 0.03 & 0.07 & -0.18 & -0.18 & 0.06 \\
\hline
\end{tabular}
\hfill
\begin{tabular}{|r|rrrrrr|}
  \hline
 & PC1 & PC2 & PC3 & PC4 & PC5 & PC6 \\
  \hline
  X29 & 0.04 & 0.04 & -0.03 & -0.07 & 0.03 & -0.00 \\ 
  X30 & -0.03 & -0.02 & 0.02 & 0.03 & -0.01 & 0.00 \\ 
  X31 & 0.01 & -0.05 & 0.05 & 0.08 & 0.02 & 0.02 \\ 
  X32 & -0.03 & 0.06 & -0.06 & -0.10 & -0.02 & -0.03 \\ 
  X33 & 0.05 & 0.06 & 0.04 & -0.07 & -0.04 & 0.22 \\ 
  X34 & -0.09 & -0.05 & -0.00 & 0.07 & 0.07 & -0.23 \\ 
  X35 & 0.00 & -0.00 & 0.00 & -0.09 & -0.13 & 0.10 \\ 
  X36 & -0.01 & 0.03 & 0.00 & 0.13 & 0.18 & -0.12 \\ 
  X37 & -0.07 & -0.11 & 0.07 & 0.04 & -0.16 & 0.04 \\ 
  X38 & 0.04 & 0.06 & -0.04 & -0.03 & 0.10 & -0.04 \\ 
  X39 & -0.08 & 0.00 & -0.00 & 0.14 & 0.09 & 0.05 \\ 
  X40 & -0.11 & 0.06 & -0.09 & -0.05 & -0.03 & -0.04 \\ 
  X41 & -0.26 & -0.31 & 0.03 & 0.10 & 0.03 & -0.12 \\ 
  X42 & 0.09 & 0.04 & 0.07 & -0.13 & -0.04 & 0.01 \\ 
  X43 & -0.12 & 0.08 & -0.04 & 0.02 & -0.15 & -0.29 \\ 
  X44 & -0.03 & -0.00 & -0.09 & 0.08 & 0.22 & 0.32 \\ 
  X45 & -0.13 & -0.04 & 0.10 & 0.07 & -0.11 & 0.04 \\ 
  X46 & -0.11 & 0.09 & -0.25 & -0.07 & 0.11 & -0.09 \\ 
  X47 & -0.29 & 0.16 & 0.12 & -0.05 & -0.18 & -0.02 \\ 
  X48 & -0.15 & 0.10 & 0.26 & 0.11 & 0.20 & -0.09 \\ 
  X49 & -0.22 & 0.10 & -0.03 & -0.01 & 0.04 & -0.06 \\ 
  X50 & -0.15 & 0.10 & -0.09 & 0.02 & -0.00 & -0.03 \\ 
  X51 & -0.22 & 0.33 & 0.07 & 0.02 & -0.14 & 0.20 \\ 
  X52 & 0.01 & 0.36 & 0.11 & 0.18 & 0.48 & -0.13 \\ 
  X53 & -0.27 & 0.14 & -0.28 & -0.01 & 0.02 & 0.04 \\ 
  X54 & -0.13 & 0.21 & -0.12 & 0.27 & -0.05 & 0.15 \\ 
  X55 & -0.18 & 0.02 & 0.22 & -0.11 & 0.00 & -0.01 \\ 
  X56 & 0.19 & -0.05 & -0.27 & 0.12 & 0.01 & -0.02 \\ 
   \hline
\end{tabular}
\caption{Same but for PCA2.}
\label{tab:PCA2}
\end{table}

\begin{table}[t]
\centering
\begin{tabular}{|r|rrrr|}
  \hline
 & PC1 & PC2 & PC3 & PC4 \\ 
  \hline
X1 & -0.07 & 0.04 & -0.02 & 0.02 \\ 
  X2 & -0.10 & -0.03 & -0.04 & -0.12 \\ 
  X3 & -0.18 & -0.13 & -0.12 & 0.16 \\ 
  X4 & 0.08 & 0.22 & 0.38 & -0.09 \\ 
  X5 & -0.06 & -0.06 & -0.13 & 0.05 \\ 
  X6 & 0.04 & 0.01 & 0.11 & -0.00 \\ 
  X7 & 0.40 & 0.18 & -0.33 & 0.12 \\ 
  X8 & 0.04 & -0.25 & 0.33 & 0.15 \\ 
  X9 & 0.12 & 0.06 & 0.03 & -0.05 \\ 
  X10 & -0.06 & -0.04 & -0.03 & 0.01 \\ 
  X11 & -0.03 & 0.05 & -0.05 & -0.02 \\ 
  X12 & 0.01 & -0.06 & 0.03 & 0.00 \\ 
  X13 & -0.06 & 0.02 & -0.06 & -0.06 \\ 
  X14 & 0.05 & -0.01 & 0.09 & 0.06 \\ 
  X15 & -0.09 & 0.01 & -0.07 & -0.03 \\ 
  X16 & 0.10 & 0.00 & 0.07 & 0.03 \\ 
  X17 & -0.03 & -0.03 & 0.03 & -0.01 \\ 
  X18 & 0.05 & 0.03 & -0.02 & 0.02 \\ 
  X19 & 0.09 & -0.10 & 0.11 & 0.33 \\ 
  X20 & -0.06 & 0.07 & -0.08 & -0.34 \\ 
  X21 & 0.21 & -0.05 & 0.00 & 0.21 \\ 
  X22 & -0.15 & 0.04 & 0.01 & -0.17 \\ 
  X23 & -0.21 & -0.04 & 0.09 & -0.14 \\ 
  X24 & 0.27 & 0.03 & -0.10 & 0.19 \\ 
  X25 & -0.30 & 0.18 & -0.05 & 0.13 \\ 
  X26 & 0.38 & -0.19 & 0.11 & -0.08 \\ 
  X27 & 0.05 & -0.21 & 0.05 & -0.07 \\ 
  X28 & -0.05 & 0.20 & -0.04 & 0.08 \\
\hline
\end{tabular}
\begin{tabular}{|r|rrrr|}
  \hline
 & PC1 & PC2 & PC3 & PC4 \\
  \hline
  X29 & -0.08 & 0.01 & 0.01 & 0.03 \\ 
  X30 & 0.04 & -0.00 & -0.00 & -0.01 \\ 
  X31 & 0.06 & -0.06 & 0.03 & -0.06 \\ 
  X32 & -0.07 & 0.06 & -0.03 & 0.07 \\ 
  X33 & -0.06 & 0.09 & 0.06 & 0.15 \\ 
  X34 & 0.08 & -0.12 & -0.06 & -0.14 \\ 
  X35 & -0.02 & 0.17 & -0.02 & 0.06 \\ 
  X36 & 0.03 & -0.24 & 0.04 & -0.07 \\ 
  X37 & 0.13 & 0.14 & 0.01 & -0.05 \\ 
  X38 & -0.08 & -0.08 & -0.01 & 0.05 \\ 
  X39 & 0.11 & -0.14 & 0.11 & -0.13 \\ 
  X40 & -0.02 & 0.06 & -0.03 & 0.06 \\ 
  X41 & 0.37 & -0.04 & -0.10 & 0.01 \\ 
  X42 & -0.14 & 0.07 & 0.05 & 0.11 \\ 
  X43 & -0.03 & 0.05 & -0.11 & 0.17 \\ 
  X44 & 0.11 & -0.15 & 0.19 & -0.27 \\ 
  X45 & 0.11 & 0.06 & -0.01 & 0.03 \\ 
  X46 & -0.02 & -0.05 & 0.04 & -0.03 \\ 
  X47 & -0.02 & 0.06 & 0.26 & 0.15 \\ 
  X48 & 0.03 & -0.28 & -0.12 & -0.00 \\ 
  X49 & 0.04 & -0.06 & 0.01 & 0.21 \\ 
  X50 & 0.01 & -0.03 & 0.05 & -0.15 \\ 
  X51 & -0.06 & 0.02 & 0.55 & 0.20 \\ 
  X52 & -0.20 & -0.59 & -0.18 & 0.19 \\ 
  X53 & 0.07 & 0.01 & 0.10 & 0.09 \\ 
  X54 & 0.04 & -0.06 & 0.07 & -0.36 \\ 
  X55 & 0.02 & -0.02 & 0.02 & 0.06 \\ 
  X56 & -0.00 & 0.01 & -0.01 & -0.02 \\ 
   \hline
\end{tabular}
\caption{Rotation onto first four PCs found for the jet cluster.}
\label{tab:PCAjet}
\end{table}

\begin{table}[t]
\centering
\begin{tabular}{|r|rrrr|}
  \hline
 & PC1 & PC2 & PC3 & PC4 \\ 
  \hline
X1 & -0.01 & 0.03 & 0.05 & -0.06 \\ 
  X2 & 0.02 & -0.03 & -0.18 & 0.19 \\ 
  X3 & -0.02 & 0.01 & -0.04 & -0.01 \\ 
  X4 & 0.07 & 0.01 & 0.08 & 0.05 \\ 
  X5 & 0.00 & -0.02 & 0.05 & -0.03 \\ 
  X6 & 0.01 & 0.02 & -0.02 & 0.02 \\ 
  X7 & 0.00 & 0.01 & -0.05 & 0.09 \\ 
  X8 & 0.06 & -0.07 & 0.10 & -0.18 \\ 
  X9 & 0.02 & 0.01 & -0.14 & -0.25 \\ 
  X10 & -0.04 & -0.04 & 0.14 & 0.17 \\ 
  X11 & -0.05 & 0.13 & -0.06 & -0.23 \\ 
  X12 & 0.00 & -0.20 & 0.08 & 0.27 \\ 
  X13 & -0.05 & 0.03 & -0.01 & 0.08 \\ 
  X14 & 0.06 & -0.04 & 0.02 & -0.10 \\ 
  X15 & -0.01 & -0.00 & 0.03 & -0.04 \\ 
  X16 & 0.01 & 0.01 & -0.03 & 0.04 \\ 
  X17 & 0.08 & 0.02 & -0.01 & -0.09 \\ 
  X18 & -0.06 & -0.02 & 0.01 & 0.08 \\ 
  X19 & 0.09 & 0.03 & 0.03 & -0.06 \\ 
  X20 & -0.08 & -0.02 & -0.03 & 0.05 \\ 
  X21 & -0.26 & 0.05 & -0.09 & -0.23 \\ 
  X22 & 0.23 & -0.06 & 0.09 & 0.20 \\ 
  X23 & 0.09 & 0.01 & -0.25 & -0.02 \\ 
  X24 & -0.08 & -0.04 & 0.22 & 0.02 \\ 
  X25 & -0.12 & -0.06 & 0.09 & 0.06 \\ 
  X26 & 0.14 & 0.05 & -0.08 & -0.06 \\ 
  X27 & 0.06 & 0.09 & -0.24 & -0.07 \\ 
  X28 & -0.06 & -0.10 & 0.25 & 0.08 \\
\hline
\end{tabular}
\begin{tabular}{|r|rrrr|}
  \hline
 & PC1 & PC2 & PC3 & PC4 \\
  \hline
  X29 & -0.04 & -0.02 & -0.05 & -0.19 \\ 
  X30 & 0.02 & 0.01 & 0.02 & 0.09 \\ 
  X31 & 0.06 & 0.21 & 0.31 & -0.02 \\ 
  X32 & -0.07 & -0.28 & -0.42 & 0.03 \\ 
  X33 & -0.10 & 0.06 & 0.06 & 0.27 \\ 
  X34 & 0.11 & -0.05 & -0.05 & -0.27 \\ 
  X35 & 0.08 & 0.37 & -0.21 & 0.07 \\ 
  X36 & -0.10 & -0.49 & 0.26 & -0.07 \\ 
  X37 & 0.13 & -0.29 & -0.27 & 0.30 \\ 
  X38 & -0.09 & 0.21 & 0.20 & -0.22 \\ 
  X39 & 0.00 & -0.07 & 0.03 & 0.12 \\ 
  X40 & 0.01 & 0.06 & -0.02 & -0.10 \\ 
  X41 & 0.01 & 0.08 & 0.04 & 0.07 \\ 
  X42 & 0.02 & -0.16 & -0.03 & -0.16 \\ 
  X43 & 0.49 & 0.09 & 0.08 & 0.04 \\ 
  X44 & -0.50 & -0.11 & -0.04 & -0.06 \\ 
  X45 & 0.22 & -0.32 & -0.04 & -0.22 \\ 
  X46 & -0.17 & 0.24 & 0.05 & 0.16 \\ 
  X47 & 0.32 & -0.07 & 0.19 & -0.06 \\ 
  X48 & -0.03 & 0.15 & -0.19 & 0.18 \\ 
  X49 & -0.03 & 0.07 & 0.14 & 0.12 \\ 
  X50 & 0.08 & -0.04 & -0.07 & -0.13 \\ 
  X51 & 0.07 & 0.07 & -0.03 & 0.03 \\ 
  X52 & 0.12 & 0.03 & 0.00 & 0.05 \\ 
  X53 & -0.02 & -0.00 & 0.03 & 0.02 \\ 
  X54 & 0.06 & 0.03 & -0.01 & -0.00 \\ 
  X55 & -0.00 & -0.00 & -0.00 & 0.00 \\ 
  X56 & 0.00 & 0.00 & 0.00 & -0.00 \\ 
   \hline
\end{tabular}
\caption{Projection on first four PCs found when performing PCA on DY results only (IDs 201, 203 and 204).}
\label{tab:PCAdy}
\end{table}

\newpage 

\bibliography{biblio}

\providecommand{\href}[2]{#2}\begingroup\raggedright\begin{thebibliography}{10}

\bibitem{Wang:2018heo}
B.-T. Wang, T.~J. Hobbs, S.~Doyle, J.~Gao, T.-J. Hou, P.~M. Nadolsky et~al.,
  \emph{{Visualizing the sensitivity of hadronic experiments to nucleon
  structure}},  \href{https://arxiv.org/abs/1803.02777}{{\ttfamily
  1803.02777}}.

\bibitem{vanDerMaaten2008}
L.~van~der Maaten and G.~Hinton, \emph{Visualizing data using {t-SNE}},
  {\emph{Journal of Machine Learning Research} {\bfseries 9} (2008) 2579}.

\bibitem{FFT74v}
M.~Fisherkeller, J.~H. Friedman and J.~Tukey, ``{PRIM-9}: {A}n {I}nteractive
  {M}ultidimensional {D}ata {D}isplay and {A}nalysis {S}ystem.'' ASA
  Statistical Graphics Video Lending Library ( {\tt
  http://stat-graphics.org/movies/prim9.html}), 1973.

\bibitem{citeR}
{R Core Team}, \emph{R: A Language and Environment for Statistical Computing}.
\newblock R Foundation for Statistical Computing, Vienna, Austria, 2018.

\bibitem{tourr}
H.~Wickham, D.~Cook, H.~Hofmann and A.~Buja, \emph{{tourr}: An {R} package for
  exploring multivariate data with projections}, {\emph{Journal of Statistical
  Software} {\bfseries 40} (2011) 1}.

\bibitem{tfep}
TensorFlow Embedding Projector {http://projector.tensorflow.org}.

\bibitem{Pumplin:2000vx}
J.~Pumplin, D.~R. Stump and W.~K. Tung, \emph{{Multivariate fitting and the
  error matrix in global analysis of data}},
  \href{https://doi.org/10.1103/PhysRevD.65.014011}{\emph{Phys. Rev.}
  {\bfseries D65} (2001) 014011}
  [\href{https://arxiv.org/abs/hep-ph/0008191}{{\ttfamily hep-ph/0008191}}].

\bibitem{Pumplin:2001ct}
J.~Pumplin, D.~Stump, R.~Brock, D.~Casey, J.~Huston, J.~Kalk et~al.,
  \emph{{Uncertainties of predictions from parton distribution functions. 2.
  The Hessian method}},
  \href{https://doi.org/10.1103/PhysRevD.65.014013}{\emph{Phys. Rev.}
  {\bfseries D65} (2001) 014013}
  [\href{https://arxiv.org/abs/hep-ph/0101032}{{\ttfamily hep-ph/0101032}}].

\bibitem{Stump:2001gu}
D.~Stump, J.~Pumplin, R.~Brock, D.~Casey, J.~Huston, J.~Kalk et~al.,
  \emph{{Uncertainties of predictions from parton distribution functions. 1.
  The Lagrange multiplier method}},
  \href{https://doi.org/10.1103/PhysRevD.65.014012}{\emph{Phys. Rev.}
  {\bfseries D65} (2001) 014012}
  [\href{https://arxiv.org/abs/hep-ph/0101051}{{\ttfamily hep-ph/0101051}}].

\bibitem{Barger:1979js}
V.~D. Barger, W.-Y. Keung and R.~J.~N. Phillips, \emph{{On psi and Upsilon
  Production via Gluons}},
  \href{https://doi.org/10.1016/0370-2693(80)90444-X}{\emph{Phys. Lett.}
  {\bfseries 91B} (1980) 253}.

\bibitem{Hou:2016sho}
T.-J. Hou et~al., \emph{{Reconstruction of Monte Carlo replicas from Hessian
  parton distributions}},
  \href{https://doi.org/10.1007/JHEP03(2017)099}{\emph{JHEP} {\bfseries 03}
  (2017) 099} [\href{https://arxiv.org/abs/1607.06066}{{\ttfamily
  1607.06066}}].

\bibitem{Asimov:1985:GTT:2812.2906}
D.~Asimov, \emph{The grand tour: A tool for viewing multidimensional data},
  \href{https://doi.org/10.1137/0906011}{\emph{SIAM J. Sci. Stat. Comput.}
  {\bfseries 6} (1985) 128}.

\bibitem{Buja:2004}
A.~Buja, D.~Cook, D.~Asimov and C.~Hurley, \emph{14 - Computational Methods for
  High-Dimensional Rotations in Data Visualization}, vol.~24 of \emph{Handbook
  of Statistics}, pp.~391 -- 413.
\newblock Elsevier, 2005.
\newblock https://doi.org/10.1016/S0169-7161(04)24014-7.

\bibitem{Cook2008}
D.~Cook, E.-K. Lee, A.~Buja and H.~Wickham, \emph{Grand Tours, Projection
  Pursuit Guided Tours and Manual Controls}, ch.~III.2, p.~295––314.
\newblock Springer Handbooks of Computational Statistics.
\newblock Springer, 2008.

\bibitem{Hou:2016nqm}
T.-J. Hou, S.~Dulat, J.~Gao, M.~Guzzi, J.~Huston, P.~Nadolsky et~al.,
  \emph{{CTEQ-TEA parton distribution functions and HERA Run I and II combined
  data}}, \href{https://doi.org/10.1103/PhysRevD.95.034003}{\emph{Phys. Rev.}
  {\bfseries D95} (2017) 034003}
  [\href{https://arxiv.org/abs/1609.07968}{{\ttfamily 1609.07968}}].

\bibitem{talk1}
{\scshape CTEQ-TEA PDF fitting group} collaboration, J.~Huston.
  \url{https://indico.cern.ch/event/647565/contributions/2707837/attachments/1519300/2372742/huston_pdf4lhc_sept2017.pdf}.

\bibitem{talk2}
{\scshape CTEQ-TEA PDF fitting group} collaboration, J.~Gao. \url{
  https://indico.cern.ch/event/568360/contributions/2481094/attachments/1438969/2214174/
  dis-ct17-jungao.pdf}.

\bibitem{Benvenuti:1989rh}
{\scshape BCDMS} collaboration, A.~C. Benvenuti et~al., \emph{{A High
  Statistics Measurement of the Proton Structure Functions F(2) (x, Q**2) and R
  from Deep Inelastic Muon Scattering at High Q**2}},
  \href{https://doi.org/10.1016/0370-2693(89)91637-7}{\emph{Phys. Lett.}
  {\bfseries B223} (1989) 485}.

\bibitem{Benvenuti:1989fm}
{\scshape BCDMS} collaboration, A.~C. Benvenuti et~al., \emph{{A High
  Statistics Measurement of the Deuteron Structure Functions F2 (X, $Q^2$) and
  R From Deep Inelastic Muon Scattering at High $Q^2$}},
  \href{https://doi.org/10.1016/0370-2693(90)91231-Y}{\emph{Phys. Lett.}
  {\bfseries B237} (1990) 592}.

\bibitem{Arneodo:1996qe}
{\scshape New Muon} collaboration, M.~Arneodo et~al., \emph{{Measurement of the
  proton and deuteron structure functions, F2(p) and F2(d), and of the ratio
  sigma-L / sigma-T}},
  \href{https://doi.org/10.1016/S0550-3213(96)00538-X}{\emph{Nucl. Phys.}
  {\bfseries B483} (1997) 3}
  [\href{https://arxiv.org/abs/hep-ph/9610231}{{\ttfamily hep-ph/9610231}}].

\bibitem{Berge:1989hr}
J.~P. Berge et~al., \emph{{A Measurement of Differential Cross-Sections and
  Nucleon Structure Functions in Charged Current Neutrino Interactions on
  Iron}}, \href{https://doi.org/10.1007/BF01555493}{\emph{Z. Phys.} {\bfseries
  C49} (1991) 187}.

\bibitem{Yang:2000ju}
{\scshape CCFR/NuTeV} collaboration, U.-K. Yang et~al., \emph{{Measurements of
  $F_2$ and $xF^{\nu}_3 - x F^{\bar{\nu}}_3$ from CCFR $\nu_\mu-$Fe and
  $\bar{\nu}_\mu-$Fe data in a physics model independent way}},
  \href{https://doi.org/10.1103/PhysRevLett.86.2742}{\emph{Phys. Rev. Lett.}
  {\bfseries 86} (2001) 2742}
  [\href{https://arxiv.org/abs/hep-ex/0009041}{{\ttfamily hep-ex/0009041}}].

\bibitem{Seligman:1997mc}
W.~G. Seligman et~al., \emph{{Improved determination of alpha(s) from neutrino
  nucleon scattering}},
  \href{https://doi.org/10.1103/PhysRevLett.79.1213}{\emph{Phys. Rev. Lett.}
  {\bfseries 79} (1997) 1213}
  [\href{https://arxiv.org/abs/hep-ex/9701017}{{\ttfamily hep-ex/9701017}}].

\bibitem{Mason:2006qa}
D.~A. Mason, \emph{{Measurement of the strange - antistrange asymmetry at NLO
  in QCD from NuTeV dimuon data}}, Ph.D. thesis, Oregon U., 2006.
\newblock 10.2172/879078.

\bibitem{Goncharov:2001qe}
{\scshape NuTeV} collaboration, M.~Goncharov et~al., \emph{{Precise Measurement
  of Dimuon Production Cross-Sections in $\nu_{\mu}$ Fe and $\bar{\nu}_{\mu}$
  Fe Deep Inelastic Scattering at the Tevatron.}},
  \href{https://doi.org/10.1103/PhysRevD.64.112006}{\emph{Phys. Rev.}
  {\bfseries D64} (2001) 112006}
  [\href{https://arxiv.org/abs/hep-ex/0102049}{{\ttfamily hep-ex/0102049}}].

\bibitem{Aktas:2004az}
{\scshape H1} collaboration, A.~Aktas et~al., \emph{{Measurement of F2($c
  \bar{c}$) and F2($b \bar{b}$) at high $Q^{2}$ using the H1 vertex detector at
  HERA}}, \href{https://doi.org/10.1140/epjc/s2005-02154-8}{\emph{Eur. Phys.
  J.} {\bfseries C40} (2005) 349}
  [\href{https://arxiv.org/abs/hep-ex/0411046}{{\ttfamily hep-ex/0411046}}].

\bibitem{Aktas:2005iw}
{\scshape H1} collaboration, A.~Aktas et~al., \emph{{Measurement of F(2)**c
  anti-c and F(2)**b anti-b at low Q*2 and x using the H1 vertex detector at
  HERA}}, \href{https://doi.org/10.1140/epjc/s2005-02415-6}{\emph{Eur. Phys.
  J.} {\bfseries C45} (2006) 23}
  [\href{https://arxiv.org/abs/hep-ex/0507081}{{\ttfamily hep-ex/0507081}}].

\bibitem{Abramowicz:1900rp}
{\scshape ZEUS, H1} collaboration, H.~Abramowicz et~al., \emph{{Combination and
  QCD Analysis of Charm Production Cross Section Measurements in Deep-Inelastic
  ep Scattering at HERA}},
  \href{https://doi.org/10.1140/epjc/s10052-013-2311-3}{\emph{Eur. Phys. J.}
  {\bfseries C73} (2013) 2311}
  [\href{https://arxiv.org/abs/1211.1182}{{\ttfamily 1211.1182}}].

\bibitem{Abramowicz:2015mha}
{\scshape ZEUS, H1} collaboration, H.~Abramowicz et~al., \emph{{Combination of
  measurements of inclusive deep inelastic ${e^{\pm }p}$ scattering cross
  sections and QCD analysis of HERA data}},
  \href{https://doi.org/10.1140/epjc/s10052-015-3710-4}{\emph{Eur. Phys. J.}
  {\bfseries C75} (2015) 580}
  [\href{https://arxiv.org/abs/1506.06042}{{\ttfamily 1506.06042}}].

\bibitem{Collaboration:2010ry}
{\scshape H1} collaboration, F.~D. Aaron et~al., \emph{{Measurement of the
  Inclusive $e{\pm}p$ Scattering Cross Section at High Inelasticity y and of
  the Structure Function $F_L$}},
  \href{https://doi.org/10.1140/epjc/s10052-011-1579-4}{\emph{Eur. Phys. J.}
  {\bfseries C71} (2011) 1579}
  [\href{https://arxiv.org/abs/1012.4355}{{\ttfamily 1012.4355}}].

\bibitem{Moreno:1990sf}
G.~Moreno et~al., \emph{{Dimuon production in proton - copper collisions at
  $\sqrt{s}$ = 38.8-GeV}},
  \href{https://doi.org/10.1103/PhysRevD.43.2815}{\emph{Phys. Rev.} {\bfseries
  D43} (1991) 2815}.

\bibitem{Towell:2001nh}
{\scshape NuSea} collaboration, R.~S. Towell et~al., \emph{{Improved
  measurement of the anti-d / anti-u asymmetry in the nucleon sea}},
  \href{https://doi.org/10.1103/PhysRevD.64.052002}{\emph{Phys. Rev.}
  {\bfseries D64} (2001) 052002}
  [\href{https://arxiv.org/abs/hep-ex/0103030}{{\ttfamily hep-ex/0103030}}].

\bibitem{Webb:2003ps}
{\scshape NuSea} collaboration, J.~C. Webb et~al., \emph{{Absolute Drell-Yan
  dimuon cross-sections in 800 GeV / c pp and pd collisions}},
  \href{https://arxiv.org/abs/hep-ex/0302019}{{\ttfamily hep-ex/0302019}}.

\bibitem{Abe:1996us}
{\scshape CDF} collaboration, F.~Abe et~al., \emph{{Forward-backward charge
  asymmetry of electron pairs above the $Z^0$ pole}},
  \href{https://doi.org/10.1103/PhysRevLett.77.2616}{\emph{Phys. Rev. Lett.}
  {\bfseries 77} (1996) 2616}.

\bibitem{Acosta:2005ud}
{\scshape CDF} collaboration, D.~Acosta et~al., \emph{{Measurement of the
  forward-backward charge asymmetry from $W \to e \nu$ production in $p\bar{p}$
  collisions at $\sqrt{s} = 1.96$ TeV}},
  \href{https://doi.org/10.1103/PhysRevD.71.051104}{\emph{Phys. Rev.}
  {\bfseries D71} (2005) 051104}
  [\href{https://arxiv.org/abs/hep-ex/0501023}{{\ttfamily hep-ex/0501023}}].

\bibitem{Abazov:2007pm}
{\scshape D0} collaboration, V.~M. Abazov et~al., \emph{{Measurement of the
  muon charge asymmetry from $W$ boson decays}},
  \href{https://doi.org/10.1103/PhysRevD.77.011106}{\emph{Phys. Rev.}
  {\bfseries D77} (2008) 011106}
  [\href{https://arxiv.org/abs/0709.4254}{{\ttfamily 0709.4254}}].

\bibitem{Aaij:2012vn}
{\scshape LHCb} collaboration, R.~Aaij et~al., \emph{{Inclusive $W$ and $Z$
  production in the forward region at $\sqrt{s} = 7$ TeV}},
  \href{https://doi.org/10.1007/JHEP06(2012)058}{\emph{JHEP} {\bfseries 06}
  (2012) 058} [\href{https://arxiv.org/abs/1204.1620}{{\ttfamily 1204.1620}}].

\bibitem{Abazov:2006gs}
{\scshape D0} collaboration, V.~M. Abazov et~al., \emph{{Measurement of the
  ratios of the Z/gamma* + >= n jet production cross sections to the total
  inclusive Z/gamma* cross section in p anti-p collisions at s**(1/2) =
  1.96-TeV}}, \href{https://doi.org/10.1016/j.physletb.2007.10.046}{\emph{Phys.
  Lett.} {\bfseries B658} (2008) 112}
  [\href{https://arxiv.org/abs/hep-ex/0608052}{{\ttfamily hep-ex/0608052}}].

\bibitem{Aaltonen:2010zza}
{\scshape CDF} collaboration, T.~A. Aaltonen et~al., \emph{{Measurement of
  $d\sigma/dy$ of Drell-Yan $e^+e^-$ pairs in the $Z$ Mass Region from
  $p\bar{p}$ Collisions at $\sqrt{s}=1.96$ TeV}},
  \href{https://doi.org/10.1016/j.physletb.2010.06.043}{\emph{Phys. Lett.}
  {\bfseries B692} (2010) 232}
  [\href{https://arxiv.org/abs/0908.3914}{{\ttfamily 0908.3914}}].

\bibitem{Chatrchyan:2013mza}
{\scshape CMS} collaboration, S.~Chatrchyan et~al., \emph{{Measurement of the
  muon charge asymmetry in inclusive $pp \to W+X$ production at $\sqrt s =$ 7
  TeV and an improved determination of light parton distribution functions}},
  \href{https://doi.org/10.1103/PhysRevD.90.032004}{\emph{Phys. Rev.}
  {\bfseries D90} (2014) 032004}
  [\href{https://arxiv.org/abs/1312.6283}{{\ttfamily 1312.6283}}].

\bibitem{Chatrchyan:2012xt}
{\scshape CMS} collaboration, S.~Chatrchyan et~al., \emph{{Measurement of the
  electron charge asymmetry in inclusive $W$ production in $pp$ collisions at
  $\sqrt{s}=7$ TeV}},
  \href{https://doi.org/10.1103/PhysRevLett.109.111806}{\emph{Phys. Rev. Lett.}
  {\bfseries 109} (2012) 111806}
  [\href{https://arxiv.org/abs/1206.2598}{{\ttfamily 1206.2598}}].

\bibitem{Aad:2011dm}
{\scshape ATLAS} collaboration, G.~Aad et~al., \emph{{Measurement of the
  inclusive $W^\pm$ and Z/gamma cross sections in the electron and muon decay
  channels in $pp$ collisions at $\sqrt{s}=7$ TeV with the ATLAS detector}},
  \href{https://doi.org/10.1103/PhysRevD.85.072004}{\emph{Phys. Rev.}
  {\bfseries D85} (2012) 072004}
  [\href{https://arxiv.org/abs/1109.5141}{{\ttfamily 1109.5141}}].

\bibitem{D0:2014kma}
{\scshape D0} collaboration, V.~M. Abazov et~al., \emph{{Measurement of the
  electron charge asymmetry in $\boldsymbol{p\bar{p}\rightarrow W+X \rightarrow
  e\nu +X}$ decays in $\boldsymbol{p\bar{p}}$ collisions at
  $\boldsymbol{\sqrt{s}=1.96}$ TeV}},
  \href{https://doi.org/10.1103/PhysRevD.91.032007,
  10.1103/PhysRevD.91.079901}{\emph{Phys. Rev.} {\bfseries D91} (2015) 032007}
  [\href{https://arxiv.org/abs/1412.2862}{{\ttfamily 1412.2862}}].

\bibitem{Aaltonen:2008eq}
{\scshape CDF} collaboration, T.~Aaltonen et~al., \emph{{Measurement of the
  Inclusive Jet Cross Section at the Fermilab Tevatron p anti-p Collider Using
  a Cone-Based Jet Algorithm}},
  \href{https://doi.org/10.1103/PhysRevD.79.119902,
  10.1103/PhysRevD.78.052006}{\emph{Phys. Rev.} {\bfseries D78} (2008) 052006}
  [\href{https://arxiv.org/abs/0807.2204}{{\ttfamily 0807.2204}}].

\bibitem{Abazov:2008ae}
{\scshape D0} collaboration, V.~M. Abazov et~al., \emph{{Measurement of the
  inclusive jet cross-section in $p \bar{p}$ collisions at $s^{(1/2)}$
  =1.96-TeV}},
  \href{https://doi.org/10.1103/PhysRevLett.101.062001}{\emph{Phys. Rev. Lett.}
  {\bfseries 101} (2008) 062001}
  [\href{https://arxiv.org/abs/0802.2400}{{\ttfamily 0802.2400}}].

\bibitem{Aad:2011fc}
{\scshape ATLAS} collaboration, G.~Aad et~al., \emph{{Measurement of inclusive
  jet and dijet production in $pp$ collisions at $\sqrt{s}=7$ TeV using the
  ATLAS detector}},
  \href{https://doi.org/10.1103/PhysRevD.86.014022}{\emph{Phys. Rev.}
  {\bfseries D86} (2012) 014022}
  [\href{https://arxiv.org/abs/1112.6297}{{\ttfamily 1112.6297}}].

\bibitem{Chatrchyan:2012bja}
{\scshape CMS} collaboration, S.~Chatrchyan et~al., \emph{{Measurements of
  differential jet cross sections in proton-proton collisions at $\sqrt{s}=7$
  TeV with the CMS detector}},
  \href{https://doi.org/10.1103/PhysRevD.87.112002,
  10.1103/PhysRevD.87.119902}{\emph{Phys. Rev.} {\bfseries D87} (2013) 112002}
  [\href{https://arxiv.org/abs/1212.6660}{{\ttfamily 1212.6660}}].

\bibitem{Aaij:2015gna}
{\scshape LHCb} collaboration, R.~Aaij et~al., \emph{{Measurement of the
  forward $Z$ boson production cross-section in $pp$ collisions at $\sqrt{s}=7$
  TeV}}, \href{https://doi.org/10.1007/JHEP08(2015)039}{\emph{JHEP} {\bfseries
  08} (2015) 039} [\href{https://arxiv.org/abs/1505.07024}{{\ttfamily
  1505.07024}}].

\bibitem{Aaij:2015vua}
{\scshape LHCb} collaboration, R.~Aaij et~al., \emph{{Measurement of forward
  $\rm Z\rightarrow e^+e^-$ production at $\sqrt{s}=8$ TeV}},
  \href{https://doi.org/10.1007/JHEP05(2015)109}{\emph{JHEP} {\bfseries 05}
  (2015) 109} [\href{https://arxiv.org/abs/1503.00963}{{\ttfamily
  1503.00963}}].

\bibitem{Aad:2014xaa}
{\scshape ATLAS} collaboration, G.~Aad et~al., \emph{{Measurement of the
  $Z/\gamma^*$ boson transverse momentum distribution in $pp$ collisions at
  $\sqrt{s}$ = 7 TeV with the ATLAS detector}},
  \href{https://doi.org/10.1007/JHEP09(2014)145}{\emph{JHEP} {\bfseries 09}
  (2014) 145} [\href{https://arxiv.org/abs/1406.3660}{{\ttfamily 1406.3660}}].

\bibitem{Khachatryan:2016pev}
{\scshape CMS} collaboration, V.~Khachatryan et~al., \emph{{Measurement of the
  differential cross section and charge asymmetry for inclusive $\mathrm
  {p}\mathrm {p}\rightarrow \mathrm {W}^{\pm }+X$ production at ${\sqrt{s}} =
  8$ TeV}}, \href{https://doi.org/10.1140/epjc/s10052-016-4293-4}{\emph{Eur.
  Phys. J.} {\bfseries C76} (2016) 469}
  [\href{https://arxiv.org/abs/1603.01803}{{\ttfamily 1603.01803}}].

\bibitem{Aaij:2015zlq}
{\scshape LHCb} collaboration, R.~Aaij et~al., \emph{{Measurement of forward W
  and Z boson production in $pp$ collisions at $ \sqrt{s}=8 $ TeV}},
  \href{https://doi.org/10.1007/JHEP01(2016)155}{\emph{JHEP} {\bfseries 01}
  (2016) 155} [\href{https://arxiv.org/abs/1511.08039}{{\ttfamily
  1511.08039}}].

\bibitem{Aad:2016zzw}
{\scshape ATLAS} collaboration, G.~Aad et~al., \emph{{Measurement of the
  double-differential high-mass Drell-Yan cross section in pp collisions at $
  \sqrt{s}=8 $ TeV with the ATLAS detector}},
  \href{https://doi.org/10.1007/JHEP08(2016)009}{\emph{JHEP} {\bfseries 08}
  (2016) 009} [\href{https://arxiv.org/abs/1606.01736}{{\ttfamily
  1606.01736}}].

\bibitem{Aad:2015auj}
{\scshape ATLAS} collaboration, G.~Aad et~al., \emph{{Measurement of the
  transverse momentum and $\phi ^*_{\eta }$ distributions of Drell–Yan lepton
  pairs in proton–proton collisions at $\sqrt{s}=8$ TeV with the ATLAS
  detector}}, \href{https://doi.org/10.1140/epjc/s10052-016-4070-4}{\emph{Eur.
  Phys. J.} {\bfseries C76} (2016) 291}
  [\href{https://arxiv.org/abs/1512.02192}{{\ttfamily 1512.02192}}].

\bibitem{Chatrchyan:2014gia}
{\scshape CMS} collaboration, S.~Chatrchyan et~al., \emph{{Measurement of the
  ratio of inclusive jet cross sections using the anti-$k_T$ algorithm with
  radius parameters R=0.5 and 0.7 in pp collisions at $\sqrt{s}=7$ TeV}},
  \href{https://doi.org/10.1103/PhysRevD.90.072006}{\emph{Phys. Rev.}
  {\bfseries D90} (2014) 072006}
  [\href{https://arxiv.org/abs/1406.0324}{{\ttfamily 1406.0324}}].

\bibitem{Aad:2014vwa}
{\scshape ATLAS} collaboration, G.~Aad et~al., \emph{{Measurement of the
  inclusive jet cross-section in proton-proton collisions at $\sqrt{s}=7$ TeV
  using 4.5 fb-1 of data with the ATLAS detector}},
  \href{https://doi.org/10.1007/JHEP02(2015)153,
  10.1007/JHEP09(2015)141}{\emph{JHEP} {\bfseries 02} (2015) 153}
  [\href{https://arxiv.org/abs/1410.8857}{{\ttfamily 1410.8857}}].

\bibitem{Khachatryan:2016mlc}
{\scshape CMS} collaboration, V.~Khachatryan et~al., \emph{{Measurement and QCD
  analysis of double-differential inclusive jet cross sections in pp collisions
  at $ \sqrt{s}=8 $ TeV and cross section ratios to 2.76 and 7 TeV}},
  \href{https://doi.org/10.1007/JHEP03(2017)156}{\emph{JHEP} {\bfseries 03}
  (2017) 156} [\href{https://arxiv.org/abs/1609.05331}{{\ttfamily
  1609.05331}}].

\bibitem{Aad:2015mbv}
{\scshape ATLAS} collaboration, G.~Aad et~al., \emph{{Measurements of top-quark
  pair differential cross-sections in the lepton+jets channel in $pp$
  collisions at $\sqrt{s}=8$ TeV using the ATLAS detector}},
  \href{https://doi.org/10.1140/epjc/s10052-016-4366-4}{\emph{Eur. Phys. J.}
  {\bfseries C76} (2016) 538}
  [\href{https://arxiv.org/abs/1511.04716}{{\ttfamily 1511.04716}}].

\end{thebibliography}\endgroup

\end{document}